\documentstyle[12pt,aaspp4]{article}

\def\teff{T$_{\rm eff}$}
\def\logg{log g}
\def\etal{et\,al.\,}
\def\kms{km\,s$^{-1}$}

\begin{document}


\title{VLT/UVES Abundances in Four Nearby Dwarf Spheroidal Galaxies:
I.  Nucleosynthesis and Abundance Ratios \footnote
{Based on Ultraviolet-Visual Echelle Spectrograph observations collected 
at the European Southern Observatory, Paranal, Chile, within the
observing programs 65.N-0378 and 66.B-0320} }


\author{Matthew Shetrone}
\affil{University of Texas, McDonald Observatory, HC75 Box 1337-L 
    Fort Davis, TX, 79734}
%\email{shetrone@astro.as.utexas.edu}
%
\author{Kim A. Venn}
\affil{Macalester College, 1600 Grand Avenue, Saint Paul, MN, 55105 \\
       University of Minnesota, 116 Church Street S.E., Minneapolis, MN, 55455}
\author{Eline Tolstoy}
\affil{Kapteyn Institute, University of Groningen, PO Box 800, 9700AV 
Groningen, the Netherlands}
\author{Francesca Primas}
\affil{European Southern Observatory, Karl-Schwarzschild Strasse 2, 85748 Garching, Germany}
\author{Vanessa Hill}
\affil{Observatoire de Paris-Meudon, GEPI, 2 pl. Jules Janssen, 92195 Meudon Cedex, France} 
\author{Andreas Kaufer}
\affil{European Southern Observatory, Alonso de Cordova 3107, Santiago 19, Chile}

\begin{abstract}

We have used the Ultra-Violet Echelle Spectrograph (UVES) on Kueyen (UT2)
of the VLT to take spectra of 15 individual red giants in the Sculptor, 
Fornax, Carina and Leo I dwarf spheroidal galaxies (dSph).
We measure the abundances of alpha, iron peak, first s-process, second
s-process and r-process elements.
No dSph giants in our sample show the deep mixing abundance pattern (O
and sometimes Mg depleted while Na and Al are enhanced) seen in nearly 
all globular clusters. At a given metallicity 
the dSph giants exhibit lower [el/Fe] abundance ratios 
for the alpha elements than stars in the Galactic halo.    
The low alpha abundances at low metallicities can be caused by a slow star 
formation rate and contribution from Type Ia SN, and/or a small 
star formation event (low total mass) and mass dependent Type II SN yields. 
In addition, Leo I and Sculptor
exhibit a declining even-Z [el/Fe] pattern with increasing
metallicity, while Fornax exhibits no significant slope.  
In contrast, Carina shows
a large spread in the even-Z abundance pattern, even over small metallicity
ranges, as might be expected from a bursting star formation history.

The metal-poor stars in these dSph galaxies ([Fe/H] $< -1$) have
halo-like s\&r-process abundances, but not every dSph exhibits the
same evolution in the s\&r-process abundance pattern.  Carina, Sculptor
and Fornax show a rise in the s/r-process ratio with increasing metallicity,
evolving from a pure r-process ratio to a solar-like s\&r-process ratio.
On the other hand, Leo I, appears to show an r-process dominated ratio over the
range in metallicities sampled.    At present, we attribute these
differences in the star formation histories of these galaxies.

Comparison of the dSph abundances with those of the halo reveals 
some consistencies with
the Galactic halo.    In particular, Nissen \& Shuster (1997)
found that their metal-rich, high R$_{max}$ high z$_{max}$ halo stars 
exhibited low [alpha/Fe], [Na/Fe] and [Ni/Fe] abundance ratios.   In
the same abundance range our dSph exhibit the same abundance pattern 
supporting their suggestions that disrupted dSph's may explain up to 50\% of 
the metal-rich halo.    Unfortunately, similar comparisons with the 
metal-poor Galactic halo have not revealed similar consistencies
suggesting that the majority of the metal-poor Galactic halo could not
have been formed from objects similar to the dSph studied here.

We use the dSph abundances to place new constraints on the nucleosynthetic
origins of several elements.
We attribute differences in the evolution of [Y/Fe] in the dSph
stars versus the halo stars to a very weak AGB or SN Ia yield of Y
(especially compared to Ba).    That a lower and flatter Ba/Y ratio
is seen in the halo is most likely due to the pattern being erased by
the large metallicity dispersion in the halo.
Also, we find [Cu/Fe] and [Mn/Fe] are flat and halo-like over 
the metallicity city range $-2 <$ [Fe/H]$ < -1.2$, and that the [Cu/alpha] 
ratios are flat.   Combining these abundances with knowledge of the 
age spread in these galaxies suggests that SN Ia are not the main site for 
the production of Cu (and Mn) in very metal-poor stars.
We suggest that metallicity dependent 
SN yields may be more promising.

\end{abstract}

\keywords{galaxies: abundances, dwarf galaxies, individual (Sculptor, Fornax,
Carina, Leo I) stars: abundances}

\section{Introduction}

Hierarchical structure formation models predict that
massive galaxies formed through continuous accretion of numerous satellites,
a process that, at a lower rate, should be continuing until today.  
One testable prediction is that the Galactic halo should have been 
formed through many minor merger events.  Another is the number of low-mass 
satellites that should be observable today around the Galaxy
(White \& Rees 1978, Moore \etal 1999, Klypin \etal 1999).
Indeed both the Galaxy and M31 contain at least one clear
remnant of a dwarf galaxy accretion event:  The tidal debris of the
Sagittarius dwarf spheroidal (dSph) galaxy (Ibata \etal 1994)
and a giant stream of metal-rich stars within the halo of M31 
(Ibata \etal 2001).  Less pronounced streams are more difficult to
detect but may stand out kinematically and in terms of abundances 
(e.g., Helmi \etal 1999).  It has also been suggested that the 
outer halo globular clusters with their predominantly red horizontal
branches did not originally form in the Galaxy but were accreted from 
dwarf satellites (e.g., van den Bergh 2000).

Thus how did the Galactic halo form, and what role did the accretion of dSph
galaxies play?  If we consider ages, dSphs can plausibly have contributed
significantly to the build-up of the Galactic halo, since the ages of
their oldest detectable populations have been found to be indistinguishable
from the oldest halo globular clusters within the measurement accuracy.
An alternative approach is to accurately {\it measure} the dSph
chemical evolution, as preserved in stellar heavy element abundance
patterns, and compare that with the Galactic halo chemical evolution.
This has been done for only a small samples of stars in a few nearby
dSphs.  The chemical evolution picture presented by Shetrone et al. (2001) 
is that the metal-poor
giants among the smallest dSphs (Draco, Ursa Minor and Sextans) have an
abundance pattern that is NOT consistent with that found in the majority 
of Galactic halo stars.   

Dwarf spheroidal galaxies can also contribute to our understanding
of the nucleosynthesis of the elements.   The difference in their
star formation histories and environments allow us to de-couple and
test some of the assumptions made in interpreting the Galactic halo
abundance patterns.   For example, the formation of even Z elements
and r-process elements are assumed to occur in SN II while the s-process
is thought to originate in AGB stars and iron peak elements from SN Ia.   
If the star formation rate, and hence the chemical evolution, is slower
in dSph then we should see a larger effect of metal-poor SN Ia and 
AGB stars than would be seen in the Galactic halo abundance patterns.   
In addition, because of the isolation of the dSph environment we can test
closed box models of chemical evolution and look for the affects of star
formation bursts and a slow star formation.  
For example, examination
of the formation of first and second peak s-process elements (e.g.,
Y/Ba) are hampered in the halo because of its mixed metallicity population
(e.g., see McWilliam 1997).    Chemical evolution in the halo occurred
very rapidly and by the time AGB stars begin to contribute to 
the ISM in the Galactic Halo there is a broad range of metallicities 
($ -3 <$ [M/H] $< -1$ ) in those AGB stars.    Studying Ba and Y 
abundances in
different environments can reveal new constraints on those
elements nucleosynthetic origins.
As another example of constraining nucleosynthetic origins of different
elements, Cu and Mn have been thought to be primarily
produced in SN Ia since Cu and Mn in the Galactic halo stars mirror the 
alpha-element abundances (Matteucci \etal 1993, Samland 1998, 
Nakamura \etal 1999), and yet other sources for
Cu have been discussed in the literature (e.g., Timmes \etal 1995).
Thus, in the halo stars, it is virtually impossible to 
distinguish SN Ia, from AGB, from metallicity dependent SN II 
nucleosynthetic sources, whereas it may be possible to disentangle
these sources with dSph abundance patterns.

In this paper, we sample four southern dSph galaxies that have not 
been previously examined: Carina, Fornax, Sculptor, and Leo\,I.    
Sculptor has a mean age similar to that of a Galactic globular cluster,
but that there was probably a spread in age of at least 4 Gyr (e.g. Monkiewicz
\etal 1999).   From low resolution spectra Tolstoy \etal 2001 found 
that Sculptor's mean metallicity was $<$[Fe/H]$> = -1.5$ with a 
0.9 dex metallicity spread.    Fornax appears to have a highly variable
star formation history spanning from $\sim$ 15 Gyr to 0.5 Gyr ago (e.g. Buonanno
\etal 1999).  From low resolution spectra Tolstoy \etal 2001 found 
that Fornax's mean metallicity was $<$[Fe/H]$> = -1.0$ with a 
1.0 dex metallicity spread.   Carina exhibits a significant variation 
in star formation rate with time with the bulk of the stars having formed
4-7 Gyr ago (e.g. Hurley-Keller \etal 1998, Dolphin 2002).
>From low resolution spectra Da Costa 1994 found 
that Fornax's mean metallicity was $<$[Fe/H]$> = -1.9$ with a 
0.1 dex metallicity spread.   Leo\,I exhibits a significant spread in 
age with the bulk of the stars having formed 2-7 Gyr ago (e.g.
Gallart \etal 1999, Dolphin 2002).
No low resolution abundance information is available for Leo\,I.    
The previous
high resolution surveys Shetrone \etal (1998, 2001) sampled Ursa Minor,
Draco and Sextans which have star formation histories similar to 
Sculptor's, dominated by a single old population.    
Comparing abundances in dSph with extremely different 
star formation histories, as well
as differences from the Galactic halo, allows us to further examine the 
nucleosynthetic sources for a variety of interesting elements.

\section{Observations}

Spectra of red giants in four dSph's were obtained at the 
Very Large Telescope Kueyen at Paranal, Chile, in August 2000
and January 2001 using the Ultraviolet-Visual Echelle 
Spectrograph (UVES; Dekker \etal 2000) in visitor mode (see Table 1).
The red arm of UVES with CD\#3 was centered at 580nm, and 
with a 1.0" slit,
we obtained a resolution $\sim$40000 (4.4 pixels) 
over a wavelength range of 480-680nm.  
The total integration time varied from 2--4 hours (1 hour per exposure),
depending on the brightness of the target and the sky conditions.
Monodimensional spectra were extracted with the UVES pipeline
(Ballester \etal 2000), then continuum normalized and combined
with IRAF for a S/N$\sim$30 per pixel. 

A variety of elements were detected in the spectra, 
including Fe, O, Na, Mg, Al, Ca, Sc, Ti, Cr, Ni, Y, Ba, Nd, La and Eu. 
This allowed for a comprehensive abundance analysis 
(e.g. Kraft et al. 1992, 1993).
Four red giants in clusters of known metallicity (see Table 2)
were observed as standard stars to establish the abundance scale. 
Analysis of these stars allowed us to look for zero point offsets 
and place our abundances on a standard system.

\section{Data Reduction and EW Measurement}

Radial velocities for each red giant (see Table 2) were measured 
from three metal lines (FeI 5083.35, CaI 6122.23, and BaII 6141.73)
and two Balmer lines (H$\alpha$ and H$\beta$).  
Heliocentric corrected radial velocities are listed in Table 2.    
The radial velocities were used to ascertain galaxy membership,
and all are in excellent agreement with published values
(see the references in Table 2).
 
Equivalent widths were measured 
three different ways using the IRAF task {\it splot}.  
The first strategy was an integrated flux method (Simpson's Rule), 
the second was a normal Gaussian fit, the third was using multiple 
Gaussians for lines that appeared asymmetric or blended with other 
lines.   In the latter cases, the Gaussian FHWM were forced to be 
the same for all components.  When the lines were not asymmetric, 
EWs were adopted from the integrated flux method, unless a bad pixel 
in the line profile made the Gaussian fit method preferable.
The adopted EW are reported in Tables 3 and 4.

Figure 1 shows a comparison of the EW measured here and those measured
for the GC sample from Minniti et al. 1993.    
There is no systematic trend or offset for the entire sample.   
The standard deviation of the entire sample is 11.5 m\AA,
however the differences are slightly higher at larger EWs 
which we attribute to a small error that scales with EW.   
We adopt the errors Minniti et al. use for their 
EW, 6~m\AA, as the minimum EW measurement error.
This uncertainty is shown by the dotted lines in 
the upper plot of Figure 1.
The dashed lines represent a combination of this minimum
uncertainty, plus a 10\% X EW uncertainty that is added in 
quadrature.  We will use this additional 10\% X EW uncertainty
later in our error analysis.
When each star is examined separately, there
do appear to be some systematic differences.
For example, our EWs for M55-283 tend to be slightly
lower than from Minniti et al. (1993), although still
in agreement to within 10\%.  We attribute these small
systematic differences to the choice of continuum normalization.    

\section{Oscillator Strengths}

Most of the oscillator strengths adopted in this work were taken from 
the Lick-Texas papers (e.g. Kraft et al. 1992 and Sneden et al. 1991) 
as summarized in Shetrone et al. (1998, 2001), and also from Fulbright (2000).   
These lines were selected for accurate abundances in metal-poor giants. 
Because several of the dSph giants in this paper are more metal-rich,
than additional lines were added from Edvardsson \etal (1993).  
In addition, UVES on the VLT has a larger spectral coverage 
than HIRES on Keck, which allowed us to add more lines.
Atomic data for these lines was obtained from the 
National Institute of Standards and Technology online
Atomic Database\footnote{Available at 
{$http://physics.nist.gov/cgi-bin/AtData/main\_asd$}. 
}.

\subsection{HFS lines}

Hyper-fine structure (hfs) plays a role in a number of elements analyzed in 
this work including Eu, Ba, Cu and Mn.   The parameters for the hfs were taken 
from a number of different references, as noted in Tables 3 and 4.
Hfs for Eu were taken from Lawler \etal (2001) but for consistancy we have
continued to use oscillator strength from Shetrone \etal (2001).  
Adopting the Lawler \etal (2001) oscillator strength would shift our Eu
abundances up by 0.08 dex.   Using the slightly higher solar abundance
in Lawler \etal (2001) would reduce this to 0.07 dex offset.

The hfs analysis was examined in all stars, but for weak lines
($<$ 40m\AA) of Cu, La, and Eu the hfs corrections were insignificant.
For the star with the strongest Eu line (Fnx 21, 87 m\AA) the hfs correction
was 0.23 dex, for all other stars the hfs correction is less than 0.12 dex.
For the Ba lines used in this analysis, the hfs corrections and isotope
splitting made no significant differences to the abundances, even for 
the strongest lines.   
Only for the Mn lines were the 
hfs corrections significant for all lines (EW $>$ 30 m\AA).

\section{Analysis}

Model atmospheres were taken from the computations of the MARCS code
(Gustafsson \etal 1975), and the abundance calculations were performed
using the Dec. 19, 2000 version of Sneden's (1973, MOOG) LTE line analysis
and spectrum synthesis code.  The procedures are identical to
those employed in Shetrone \etal (2001) ensuring that the relative
abundance and model parameter scales should be similar.
In general, a color temperature and metallicity were adopted
per program star (discussed below), and the initial temperature 
was adjusted to minimize the slope in Fe abundance (from Fe I) versus 
excitation potential.   Minimizing the slope between FeI line
abundances and their equivalent widths
also provided the microturbulent velocity.   
Following this, the surface gravity was determined by requiring 
that the abundance of the {\it ionized} species equal that of 
the {\it neutral} species based upon Fe I and Fe II.   
These steps usually required a few iterations before the
parameters converged and were adopted for the abundance analysis.  
Model atmospheres are from the MARCS grid that are slightly more 
metal-rich than the actual derived abundances to compensate for 
the extra electrons that are contributed by alpha-rich metal-poor stars
(see Fulbright \& Kraft 1999 for more about this methodology).
Model atmosphere parameters determined here are listed in Table~5.

In addition, we performed two checks on our model atmospheres analyses.
Firstly, the final model temperatures were examined relative to the 
initial color temperatures derived from the B-V colors.   
Secondly, the iron (and other) abundances for two stars were also
analysed using ATLAS9 model atmospheres (Kurucz 1993) in
WIDTH9 with oscillator strengths from the VALD database 
(Kupka et al. 1999).  The two tests are discussed separately below.

The B-V color for each star provided an initial estimate for the stellar
parameters.  The conversion from color to stellar parameters was made 
using a calibration based upon the derived parameters for a number of 
globular cluster stars (Lick-Texas papers; Kraft \etal 1992, 
1993, 1995, 1996, Sneden \etal 1991, 1997).      
Initial estimates were made by assuming a metallicity for each program
star based upon their location in the color magnitude diagrams, then
these estimates were adjusted for the metallicities actually determined
per star.     Because the iterative nature of our analysis
the final temperatures and surface gravities do not match the initial
estimates.     On average the temperatures differed little from the 
initial estimates ($\Delta$T $= -3$K, $\sigma = $ 92K) while the 
final surface gravities are a bit lower than
the initial estimates ($\Delta$log g $= -$0.29 dex, $\sigma =$ 0.17 dex).
Colors were taken from Schweitzer et al. 1995 (for Sculptor), 
Mateo et al. 1991
(for Fornax), Mateo et al. 1993 (for Carina), and Mateo et al. 1998 (for
Leo I).  Reddening estimates were taken from Kaluzny et al. 1995 
(for Sculptor), Schlegel, Finkbeiner \& Davis 1998 (for Fornax), 
Mould \& Aaronson 1983 (for Carina) and 
Cardelli, Clayton \& Mathis 1989 (for Leo I).   
A second check of our adopted stellar parameters was performed using the
the Alonso temperature scale (see Table 2 in Alonso et al. 1999 )
and then using that effective temperature and the new Yale-Yonsei isochrones
(Yi \etal 2001, Green \etal 1987) to derive surface gravity.    
The Alonso temperature scale (making the correction in Alonso \etal 2001) and 
literature B--V colors suggest a slightly cooler temperature than our adopted 
temperatures ($\Delta$T $=$ +60K, $\sigma = $ 107\,K), however the
surface gravities based upon the isochrones is in good agreement with 
our adopted gravities ($\Delta$log g $=$ -0.07 dex, $\sigma =$ 0.18 dex).
The Tolstoy \etal (in prep) use cousins I while the 
Alonso use Johnson I.   Using Bessell (1986,1990) to convert the colors and 
converting the 
E(B--V) to E(V--I) using Dean, Warren \& Cousins 1978 we find a similar 
zero point ($\Delta$T $=$ +51K, $\sigma = $ 131\,K)
between our derived temperatures and the Alonso temperature scale.  The 
large dispersion in the between in both V--I and B--V could be due to 
variable reddening.      
Inspection of the spectra reveals a factor of two
dispersion in EW of the interstellar Na D lines among the Carina sample.
As mentioned before we have adopted the spectroscopic temperatures
and have only used the photometric temperatures as an initial 
estimate and a secondary check on our methodology.

Two stars, the cluster star M55-76 and the Sculptor star Scl-459, 
were checked with ATLAS9/WIDTH9 calculations and VALD atomic data.
The abundances for Fe~I and Fe~II lines are in very good agreement 
($\Delta$log({\sl X}/H) $\le$ 0.1), and most of the iron line
abundance disagreements can be traced primarily to small differences 
in the oscillator strengths.
We note however that the mean differences go in opposite directions for 
FeI and FeII, so that the ATLAS/WIDTH results do not maintain the iron 
ionization equilibrium when the MARCS/MOOG parameters are adopted.
For example, when FeI=FeII using MARCS/MOOG, then the ATLAS/WIDTH/VALD 
results are FeI + 0.1 dex = FeII $-$ 0.1 dex, resulting in a 0.2 dex
difference between iron from the FeI versus the FeII lines. 
This will affect the model atmosphere parameters, primarily it will
force a higher gravity determination in the ATLAS/WIDTH analysis.   
While gravity has a very small effect on the FeI abundances 
(see Table 6), and thus the overall metallicity adopted for that model,
it can have a larger effect on the abundances of ionized species and
also the O~I abundance.  This is discussed further below in Section 6.3. 
Additionally, we stress that the MARCS/MOOG analysis is the most
consistent with the published abundances for the globular cluster 
standard stars and for red giants in other dwarf spheroidal galaxies, 
thus we consider these the most appropriate for differential comparisons.

\section{Error Analysis}

We divide our errors into three types: statistical, internal and external.
Statistical uncertainties are those errors which can be reduced by using many
lines to measure the abundances.   The internal errors are those errors
based on analysis methodology, such as derivation of T$_{eff}$ or normalization
of the continuum.    The external errors are those based upon the analysis
tools, such as the model atmosphere grid and LTE abundance analysis code.

\subsection{Statistical Errors}

The statistical errors are determined from the consistency of the abundances
derived from each line.    Assuming that our derived stellar parameters
are approximately correct, the variance in the abundance derived for elements 
with many lines, such as Fe I, is a measure of our ability to measure 
consistent EWs and the accuracy of our atomic physics inputs (largely
the oscillator strengths and hyperfine structure).    
Using the Cayrel formalism (1988), we estimate that our random error 
in EW should be 4m\AA\ for the dSph stars.   The Cayrel formalism simply
assumes a line profile affected simply by the S/N (30 in our case), and
the number of pixels in the resolution element (4 pixels and R=40,000).  
For the weakest lines 
( $\sim$ 10 m\AA), this will introduce an uncertainty of 0.19 dex.   
For moderately strong lines ( $\sim$ 60 m\AA), the uncertainty is 0.03 dex,
while for very strong lines ($\sim$ 150 m\AA), its only 0.01 dex.  
The globular cluster spectra have much higher S/N, thus they will also
have smaller EW errors.   As mentioned earlier in our comparison
of our EW to the Minniti et al. (1993) EW our errors were better
represented by a constant with a 10\% X EW additional error.   Thus
we take our actual error to be 4m\AA + (10\% X EW), in the case 
for the dSph sample).   
 
Since many elemental abundances are derived from only a few lines, 
then the statistical error is rarely accurately sampled.  
Thus, we assume that the standard deviation of the Fe~I 
line abundances is typical for most elements.
We will refer to this $\sigma$ as the {\it average line deviation}.
For each element we take the larger of either  
(a) the standard deviation of the mean  of the lines for that element, 
assuming that there is more than 1 line,  (b) the average line deviation 
divided by the square-root of the number of lines used to determine the 
abundance for that element, or (c) for elements with only 1 line, then
the error based just on EW using the Cayrel formalism plus the (10\% X EW)
additional error we described earlier.

In Tables 8-11 we have given the abundances and internal statistical error
for each element.  For Fe I we have listed the number of lines that 
went into the calculation of the standard deviation of the mean.   For
the other elements a letter tag is given which represents which 
method is used.   
An "S" means that the standard deviation was taken from that element.
An "I" means that the average line deviation method was used.   
An "E" means that the error is derived from the EW error.
No uncertainty is given if only an upper limit to the abundance is determined.

\subsection{Internal Errors}

In Section 5, we computed the difference between our derived
stellar parameters and those based upon photometry.    From that analysis,
we adopt internal uncertainties of $\pm$100K and $\pm$0.2 dex for 
\teff and \logg, respectively.   We also estimate that the error in 
the microturbulent velocity is $\pm$0.2 \kms.   
Table 6 lists these effects on the abundances for
one star, Car 2, by recomputing the abundances for models with slightly
different parameters.    We have also listed the effect of choosing
a slightly more metal-poor model (i.e., one without the extra metallicity
which compensates for the alpha-rich abundance pattern), and the
effects of shifting the continuum systematically up such that all of
the EW are 4m\AA\ larger.     This continuum error assumes that the 
line profile width does not grow significantly with EW.   This is clearly
not true for the very large EW lines but we have made some effort to 
remove all strong lines from this analysis so to first approximation this
is a reasonable assumption.  For the globular cluster stars the S/N is
much higher and thus we adopt a smaller error in the continuum.

It should
be noted that many of these errors are not independent, e.g., a 
change in the \teff\ by 100~K introduces a slope in the Fe I line
abundances vs. EW plot which is used to determine the microturbulent 
velocity.   A 100~K change in the \teff\ also upsets the balance of 
the Fe~I vs. Fe~II abundances.   The last column in Table 6 shows 
how the abundances would change if we attempted to mediate the effects 
by recomputing the abundances with a model that was 100K too cool.    
We adopt this last column as representing the most accurate 
abundance error based on changes in \teff, \logg\ and microturbulence.

To combine the uncertainties per element due to the stellar parameters,
continuum placement, and metallicity, we have taken these uncertainties
in Table 6 and combined them in quadrature over the entire range of
stellar parameters.   These {\it total internal uncertainties} 
are listed in Table 7.   

In this paper, plots of abundances will combine the 
statistical uncertainty and the internal uncertainties in quadrature
to create a single error bar.
 
\subsection{External Errors}

External errors due to model atmospheres and analysis methods can be 
extremely difficult to diagnose and quantify.    
For example, using spectral indicators to determine the stellar 
parameters rather than relying on the photometrically derived 
parameters can shift all of the \teff\ and/or \logg\ values systematically 
up or down, which will affect the abundances.  The magnitude of the 
effect on each element can be estimated from Table~6.   
As a demonstration, if the Alonso temperature scale were adopted, then
a shift in temperature by $-$60~K would have occurred, which shifts all 
of the [FeI/H] abundances down by 0.06~dex.   While this shift is small,
it would also have occurred to the globular cluster standard star results.
Since the interpretation of the abundances in the dSph galaxies depends 
on a differential comparison with the globular cluster standards, 
then these small systematic shifts would not have a significant effect
on the final results.

On the other hand, our comparison of MOOG/MARCS abundance results 
to those from ATLAS/WIDTH/VALD may be more valuable.   
As an example, the mean abundance results for the Sculptor
star Scl-459 from each analysis method are shown in Table 12.   
As discussed above, the changes to the iron ionization 
equilibrium would force a slightly higher gravity in an 
ATLAS analysis.   Small changes in gravity would have a negligible 
effect on the abundances from most of the neutral species, but a
more significant effect on the derived abundances for O~I and the
ionized species.   Thus, the absolute O/Fe ratio determined for an 
individual star could be affected (note that accurate O/Fe abundances 
is a problem with a large scope in metal-poor stars, and we refer 
to more specific papers on this problem, e.g., Lambert 2001, 
Asplund \& Garcia Perez 2001).
In this paper, the interpretation of the O/Fe ratio is 
done with respect to standard stars whose analyses are
done using the same techniques as the dSph stars.
Thus, the {\it differential} O/Fe abundance ratios are 
similar whether derived from a MARCS/MOOG analysis or using 
the ATLAS/WIDTH techniques.    The effect of changing the surface
gravity on the ionized species is larger. 
While all of the s-process abundances could be affected by a significant
amount (see Table 6, e.g., Ba~II/Fe, Eu~II/Fe), the comparison of 
BaII/YII or BaII/EuII will be far less affected.  In 
addition,  most of our comparisons of the ionized species abundances, 
such as [BaII/Fe], should be similarly unaffected if the affect is systematic
since our comparisons will be made between our 
globular cluster giants and our dSph giants.

Other comparisons of abundance results in Table 12 show that there
are no further changes between the analyses techniques by greater
than 0.1~dex (the hfs of Mn and Cu were not included in the ATLAS/WIDTH
analysis).    It is also interesting to note that differences in the
gf-values can still be important (causing $>$0.1 dex differences)
in the analyses of Al, Sc, and Ti. 

No corrections have been made for non-LTE effects on our abundances have
have attempted to compare our abundances with similiar LTE analyses 
to minimize this source of error.  

\section{Globular Cluster Abundances}

Four red giants in three globular clusters were observed as
standard stars to check our data reduction and analysis methods.    
There is excellent agreement in the metallicities derived in this 
paper with the iron abundances from Minniti et al. (1993), 
where $\delta$[Fe/H] $= -0.03$ with $\sigma = $ 0.16 dex,
despite different line sets and oscillator strengths.   

The globular cluster stellar abundance ratios are shown in Table 8. 
The abundances for these globular cluster stars are typical
of those published for the halo (c.f., McWilliam 1997) to within
the statistical and internal errors, with the exception of Ti. 
Our Ti abundances fall about 0.15 dex below the typical Ti 
abundances, e.g., the Ti abundances from Fulbright (2002) who
used the same line lists and very similar methodology.   
Reanalysis of the Shetrone \etal (2001) dSph and globular cluster 
spectra using only the lines adopted in this analysis revealed only 
slightly smaller abundances (0.05 dex).   Thus, we can not account 
for this discrepancy, and will limit our discussion of Ti in the
dSph stars to differential abundances only.

We find that two, possibly three, of our four globular cluster 
standard stars show deep mixing.   For metal-poor stars 
(with [Fe/H] = -2.0), deep mixing is detected as a star 
showing high [Al/Fe] and [Na/Fe],
but low [O/Fe] and possibly low [Mg/Fe] (Shetrone 1996).    
In our sample, M30-D, M55-283, and M68-53 exhibit abundance
ratios consistent with this pattern (see Figure 2).   
Only M55-76 does not appear to have undergone deep mixing.
For the Galactic field halo stars, the [O/Fe] and [Mg/Fe] 
abundances can be grouped with the other even Z elements 
when there is no evidence of the deep mixing pattern,

\section{Dwarf Spheroidal Abundances}

In this paper, we discuss the abundance pattern in the dwarf
spheroidal stars by element and discuss the nucleosynthesis 
of these elements in comparison to the Galactic halo.   A
discussion of the element ratios by galaxy can be found in
Tolstoy \etal (2002, hereafter Paper II).   
Only Carina will be discussed separately
here, which may show an 
alpha-element abundance pattern consistent with theoretical 
predictions for its bursting star formation history. 

\subsection{No Deep Mixing in dSph Stars}

The surface abundances of Al and Na are very sensitive to
deep mixing in red giants.   Two (possibly three) of our
globular cluster standards show elevated Al in Figure 2.
In contrast, all of the dSph stars have halo-like Al/Fe ratios.  
One object in LeoI (Leo-5) may show a slightly elevated abundance 
([AlI/FeI]=+0.42), although this star shows a normal field halo-like
Na and O abundances.  In fact, all dSph stars show 
sub-solar [Na/Fe] ratios.
Thus, we do not expect any of the dSph stars have undergone 
deep mixing.  As such we will include O and Mg in our discussion 
of the even Z elements.    The Na abundances in our study 
are consistant with the Stephens (1999) study of halo Na but
fall below other studies including our globular cluster
sample, Gratton \& Sneden 1988 and McWilliam 1995.  The 
Stephens 1999 sample were selected to probe the outer halo
and thus may be a slightly different sample than the other
halo studies.  This will be discussed in a later section.

\subsection{Even Z Elements}

The theoretical picture for the formation
of even-Z elements (O, Mg, Si, Ca, Ti)
is in the nucleosynthetic shell-burning
during SN II at the end of the life of
massive stars.    This hypothesis is supported by 
elemental abundances in halo stars (c.f., McWilliam 1997).   
It is also important to note that this theoretical picture
generally applies to elements formed by alpha-capture, but
the results from the halo stars suggest that Ca and Ti also follow
this predicted behavior, and Ca and Ti are therefore lumped in with
the alpha-elements.   We will make a subtle distinction between 
the true (easy to understand) alpha elements, O, Mg and Si from
the heavy even-Z elements Ca and Ti.   

In the canonical picture of Galactic halo formation 
the even-Z elements are produced 
en masse shortly after a burst of star formation with so little 
time elapsing that SN Ia have no time to dilute the pure SN II 
abundance pattern.   
At later epochs ($>$1.0 GYR) SN Ia had a chance to contribute.   
SN Ia are thought to 
produce little to no O and Mg while they probably
are able to produce significant amounts  
of the iron peak even-Z elements Si, Ca and Ti (see Woosley \& Weaver 1995 and 
Table 3 in Iwamoto \etal 1999).    
Because of the under production of 
O and Mg by SN Ia  (Iwamoto \etal 1999) the [O/H] and [Mg/H] 
should remain constant 
and the [O/Fe] and [Mg/Fe] abundance ratios should decrease
with increasing metallicity.    
Because SN Ia produce some Si, Ca and Ti
but less than are produced in SN II, 
the [Si/H], [Ca/H] and [Ti/H] will rise slightly and the 
[Si/Fe], [Ca/Fe] and [Ti/Fe] will decrease slightly.  In this
scenario the even-Z elements slowly transition from a high value to a
solar value with increasing metallicity (time).

The yields of the alpha abundances
with respect to iron abundances in SN II  are mass dependent 
(Woosley \& Weaver 1995) with higher masses producing a larger
percentage of alpha elements with respect to iron.   If a small
star formation event occurs, where relatively few high mass stars
are formed, then the most massive SN II may not be present and
the ratio of alphas to iron could be altered from 
the canonical halo SN II abundance pattern.  For example, with 
a Salpeter IMF (Salpeter 1955, and also see Massey 1998 for the 
IMF for the Local Group) and a small 1000 solar mass star formation 
event, it is statistically unlikely that stars over 25 solar masses 
will form.    Using the Woosley \& Weaver 1995 SN yields
such an event will have much lower [O/Fe], [Mg/Fe] and [Si/Fe] 
abundance ratios (by 0.4 to 0.6 dex) with respect 
to a much more massive star formation event where many higher mass
stars are likely to form.   
This was also noted by Gibson (1998) in an examination of 
the upper limit to the IMF.
{\it Thus,
a low mass star formation event could produce abundances that are slightly
less enhanced than those found in the halo.   }

As shown in Figures 3 to 8, even-Z abundance ratios are 
generally larger than solar in our metal-poor stars, as also
seen in the halo.     
To produce these ratios
requires reasonably massive early star formation events.
The most metal-poor star in the Sculptor sample (H400) and
the more metal-poor star in Leo I (Leo 5) have 
alpha element ratios consistent with that of the halo 
(Gratton \& Sneden 1988,
Gratton \& Sneden 1991, Gratton \& Sneden 1994, McWilliam \etal 1995, 
Stephens 1999) 
indicating only a minor (if any) contribution from SN Ia.   
In Sculptor (Figures 4 and 7) and LeoI (Figures 5 and 8), 
the even-Z to iron ratios appear to decrease as Fe increases.    
These trends are based upon few data points and thus should 
be viewed carefully.
To produce the decline in the even-Z abundance ratios requires either 
a later epoch of SN Ia contributions, or
a later stage of small star formation events which had fewer 
high mass SN II and thus produced lower
even-Z abundance ratios.\footnote{For a review on the 
star formation histories
of the Local Group galaxies, see Mateo 1998.   A more
detailed discussion of the star formation histories of
our four dSph galaxies is included in Paper II.}.    

In Fornax (Figures 5 and 8), the even-Z ratios appear flat to 
slightly rising.   
The average of the [O/Fe], [Mg/Fe] and [Si/Fe] abundance ratios 
is 0.1 dex ($\sigma 0.1$) which is significantly smaller than 
that of the halo and our globular cluster sample 
(excluding the Mg and O for the stars with the deep mixing pattern). 
Again this can be done either through SN Ia contributions or 
from later smaller star formation events (lower mass SN II contributions).
Fornax has a large spread in ages as indicated from its color-magnitude diagram
(c.f., Mateo 1998, Paper II).    Thus, one expects to have 
significant contribution from SN Ia in the younger (more metal-rich)
population.    

The alpha-element abundance pattern in Carina (Figures 3 and 6)
exhibits a large and interesting dispersion that we will address 
separately below.

\subsection{Fe-peak Elements}

The Cr, Co, and Ni abundances in the dSph stars are halo-like
(Gratton \& Sneden 1988, 1991, 1994,
Sneden \etal 1991, 
McWilliam \etal 1995, 
Stephens 1999) 
i.e., they remain constant with FeI to within the errors down to 
[Fe/H]$\sim -$2, as seen in Figure 9.   
Two stars near [Fe/H]$\sim -$1.1 (LeoI-2 and Scl-H482)
may also show slight Ni underabundance.   This 
is interesting because Nissen \& Shuster (1997) found a 
puzzling relationship between Ni and Na (and alpha-elements) 
in this same metallicity regime in halo stars;   
a tiny decrease in Ni is accompanied by a moderate 
decrease in Na (and alpha's) near [Fe/H]$\sim -$1.  
The Ni underabundance also seems to be related to lower
alpha-abundances (and possibly Na) in these two dwarf spheroidal
stars.   

Sc is also halo-like (i.e. flat near 0.0 dex) 
for most of our targets, however a few 
stars (Leo 2, Car 3 and Scl-482) have significant underabundances.   
We also notice that the Sc abundances plotted in Figure 9 mimic
the pattern of the alpha-elements better than that of the iron-group
elements.  Because the nucleosynthetic origin for Sc is unclear we 
will not comment further on Sc.

The Zn abundances in our dSph sample are systematically a few dex 
lower than those found in the Galactic halo (Sneden \etal 1991, Primas
\etal 2000) and
in our globular cluster sample.     This seems to imply that the Zn is
behaving differently from the other iron peak elements in ALL of 
these dSph.   This is not entirely surprising since the nucleosynthetic
origin of Zn is uncertain 
with possible origins in SN Ia, SN II and/or AGB stars
(Matteucci \etal 1993, Hofman \etal 1996, Umeda \& Nomoto 2002).

\subsection{Cu \& Mn}

The formation sites for Cu and Mn are not well known.
In halo stars, the Cu and Mn ratios are both less than 
solar until [Fe/H] $\sim -$1.0 when they both rise to solar
Gratton \& Sneden 1988, Gratton 1989, Sneden \etal 1991,
McWilliam \etal 1995).   
The most common interpretation of this pattern is that 
they are produced in SN Ia (Gratton 1989, 
Matteucci \etal 1993, Samland 1998, Nakamura \etal 1999).   
Alternatively, Woosley \& Weaver (1995) have
suggested a metal-dependent SN II yield, 
such that at [Fe/H] $\sim -$1 the metallicity 
becomes sufficiently high that significant amounts
of Cu and Mn can be produced in the SN ejecta 
(see Timmes \etal 1995 for a chemical evolution model 
using the Woosley \& Weaver 1995 yields).

As shown in Figure 10, our Cu and Mn ratios are consistent with 
the halo star abundances.   They are less than solar over a wide 
range of low metallicities up to [Fe/H] $\sim -$1.   
The similar Cu, Mn, and Fe abundance patterns between the Galactic 
halo stars and the dSph stars suggest a similar abundance origin.   
In Figure 10, we also notice that [Cu/alpha] is significantly
less than solar and flat for the dSph stars (until [Fe/H]$>-$1).
{\it This strongly suggests that either SN Ia do not
contribute to Cu in the most metal-poor stars, like the alpha-elements,
or that any SN Ia contribution to Cu at this metallicity is not significant}.
If significant amounts of Cu were being produced in metal-poor
SN Ia events, then as Fe increases we would expect Cu/alpha to increase.

ne may question then whether SN Ia products are contributing
at all up to [Fe/H]=$-$1.    As discussed in Section 8, either
SN Ia are contributing to explain the alpha/Fe ratios, or possibly
only small star formation events have occurred (thus lower mass SN II).
However, also given the star formation histories for these
galaxies, interpreted from their CMDs (see Paper II),
it would be surprising if there were no SN Ia contributions until
[Fe/H]=$-$1.   All of these galaxies are thought to have had some
star formation in the distant past (15 Gyr), with either
continuous or bursting star formation at intermediate ages (5-10 Gyr).
The intermediate-aged stars can be expected to form from gas
enriched in SN Ia products from the earlier generation(s).
Thus, we suggest that if Cu is produced in SN Ia, then the yield
may be metallicity dependent, with increasing amounts of Cu as
metallicity increases.    It is also possible that the upturn
in Cu/Fe near [Fe/H]=$-$1 is due to a metallicity dependent SN II yield.
This conclusion is not sensitive to the choice of hfs or gf values because
it is based on relative abundances within this analysis.

A similar argument can also be made for Mn.  Figure 10 shows
that Mn/Fe is also flat and halo-like.  The halo stars appear
to have increasing [Mn/Fe] above [Fe/H]=$-$1.    In the halo,
the upturn has been intrepreted as the onset of SN Ia products.
We suggest that, like Cu, SN Ia (or even SN II) contributions
may be metallicity dependent with very little Mn produced until
[Fe/H]=$-$1.

Omega Cen is another system where [Cu/Fe] is quite low
over the same range of ages and metallcities as our
dSph stars (Cunha \etal 2002).   Unlike the halo stars,
[Cu/Fe]$\sim -$0.5 in the star in Omega Cen and does not
increase with metallicity.   Our dSph results are not
inconsistent with this result either, since our Cu/Fe ratios
do not increase as quickly as in the halo stars.   Cunha
\etal similarly conclude that SN Ia contribute very little
to the chemical evolution of Cu in the metallicity range
of $-$2.0 $<$ [Fe/H] $<$ $-$0.8.  In contrast, Pancino \etal (2002)
found an increasing [Cu/Fe] abundance with increasing
metallicity for Omega Cen
giants in the metallicity range of $-$1.2 $<$ [Fe/H] $<$ $-$0.5.
Both of these results could be interpreted as pollution from SN Ia
or as metal dependent SN II yields.
(A more detailed comparison of the abundance patterns
observed in Omega Cen with those found in
the dSph is beyond the scope of this paper).

\subsection{The First s-process Peak Element, Y }

Y samples the first s-process peak, which may 
have a different source than the heavier (e.g., Ba) 
s-process elements.
In Figure 11, we notice that our most metal-poor stars,
have halo-like [Y/Fe] and [Ba/Y] ratios, implying a
similar origin or different sources in the same proportion as were 
found in the halo (e.g., early SN II  yields).   But, as the 
metallicity increases, the [Y/Fe] abundance ratios decrease.
This representation of the Y abundances is a bit misleading
though.   The central plot in Figure 11 shows the absolute 
Y abundances, [Y/H], where it can be seen that Y does actually
increases with metallicity for Carina, Leo I and Fornax.
The [Y/Fe] ratio decreases though because the Fe abundance 
is increasing more rapidly than the Y abundance is increasing
in these dSph galaxies in comparison to the Galactic halo.
For Sculptor, the [Y/H] abundance has a wide dispersion but 
remains constant over the metallicity range we sample.

A model for the formation of s-process elements in AGB stars 
by Clayton (1988) suggests that the yields scale with metallicity 
if the neutron source is the $^{13}$C($\alpha$,n)$^{16}$O reaction, 
and this model specifically predicts that [Ba/Y] should increase 
with metallicity.  The bottom panel in Figure 11 shows that [Ba/Y] 
clearly does increase with metallicity in the dSph stars as predicted.
That this pattern is not seen in the halo stars is more
peculiar, and suggests a number of possibilities.  
McWilliam (1997) discussed that the predicted [Ba/Y] relation 
in the halo may have been erased by the large metallicity 
dispersion in the halo, i.e., at any given time, the secondary 
elements are produced from stars with a variety of metallicities 
and thus yields.   This interpretation predicts that the  
rising [Ba/Y] ratio in the dSph is caused by 
chemical evolution occurring over
a longer period of time (in comparison to the halo) and thus AGB stars of 
a narrower range (in comparison to the halo) in metallcity are contributing 
to the ISM.
Another option might be that that the seed for the first 
s-process peak (C?) is underabundant in the dSph galaxies.
Low resolution spectra of several dSph's show high carbon 
abundances though with respect to Galactic globular clusters 
of similar metallicities
(Kinman \etal 1980, Smith \& Dopita 1983, Smith 1984,  Bell 1985).
A third option, if we forgo Clayton's model, could be that there
is a source of Y in the Galactic Halo that is not present in the
higher metallicity dSph stars.   Since most studies of SN II yields
do not include the first s-process peak then we cannot compare this
hypothesis with any models.    The first option above is the most
consistent with the overall abundance patterns.

The Y enrichment in the metal-rich Fornax star, Fnx-21, 
is consistent with other s-process enrichments
in this star (discussed below). 

\subsection{s-process and r-process Elements}

In the Sun, Eu is largely an r-process element, 95\% (Burris \etal 2000).
The site of the 
r-process has been suggested to be low mass SN II (Mathews \etal 1992),
but the site for the r-process is still a matter of debate 
(e.g. Wallerstein \etal 1997, Tsujimoto \& Shigeyama 2001, Qian 2002).  
However, most of these models share a common prediction: SN II
 are the source of the r-process.  Thus,
[Eu/H] should rise whenever SN II contribute to the ISM, and 
only when SN Ia and AGB stars contribute to the ISM should 
[Eu/H] remain constant and the [Eu/Fe] ratio decline.    
The Eu abundances are plotted in Figure 12.   
In Leo I and Fornax the [Eu/H] abundance increases with 
metallicity as expected if there has been some ongoing star 
formation with SN II contributing to the ISM.  
A similar slope is also seen in the Galactic halo stars 
(Gratton \& Sneden 1991,
Gratton \& Sneden 1994, McWilliam \etal 1995, Burris \etal 2000).   
Thus we predict a burst of star formation between [Fe/H] 
$-1.5$ to $-1.1$ for Leo I and $-1.5$ to $-1.2$ for Fornax.
These predictions are provisional given the few number
of points and the large errorbars.
On the other hand, the Sculptor [Eu/H] abundances are relatively
flat over the entire metallicity range sampled which implies 
little to no later contribution of SN II to the ISM.   
Thus for Sculptor, we predict that only a single burst occurred 
or that the material from SN II was completely lost from the galaxy
in any later bursts.
The Carina abundances will be discussed separately below.    

Oddly, the most metal-poor star in Sculptor, H-400, 
has a larger [Eu/H] abundance than Galactic halo 
stars of similar metallicity.    
The top panel of Figure 12 shows that this star has [Eu/Fe] $= +1.0$, and,
as we will show later, an r-process dominated abundance pattern.  This
type of super r-process rich abundance pattern has been seen among 
Galactic halo stars (McWilliam \etal 1995) 
and attributed to inhomogenous mixing
of the SN II yields (McWilliam \etal 1997), i.e., 
the star forms after the local
ISM is contaminated by a nearby r-process rich SN II and before the ISM
is well mixed.  However, all of these Galactic r-process rich stars are
more metal-poor than H-400.    This high [Eu/Fe] abundance could be due
a wide dispersion in [Eu/H] at [Fe/H] $= -2.0$ and we have only
sampled the upper end of that distribution.   

A comparison of s\&r-process elements to Eu
(a largely r-process element) allows us to
examine the contributions to the abundances 
from AGB stars.    The s-process/r-process ratios are shown in Figure 13.  
The pure r-process contributions to these elements
from Burris \etal (2000) are shown by the dotted lines,
while the solid line shows the pure r-process contributions 
from Arlandini \etal (1999).   The Burris \etal contributions
are calculated using the ''classical approach'' which models
the neutron flux of an AGB star with a simple analytical model.
The Arlandini \etal values come from a new generation of 
AGB evolutionary models.   Of course comparison to these 
solar system fractions requires our abundances to be on an absolute
scale and introduces many additional concerns.   The r-process 
fractions should be considered free parameters able to slide up 
or down within our abundance scale.
It is important to stress that La and Ba
are often called s-process elements based upon the fraction of these
elements that were produced by the s-process {\bf in the Sun}.  However, 
in the early Universe and apparently in the most metal-poor stars in these
dSph we expect that all the heavy elements present have their origins in 
the r-process since AGB stars would not have had time to evolve and 
contribute to the ISM (Truran 1981).
Indeed, in our most metal-poor stars, the Ba, Nd and La abundances are 
consistent with primarily r-process contributions.
Note that in the Sun, La and Ba are mostly s-process elements 
(85\% and 75\% respectively from Burris \etal (2000) while 
Nd is thought to be (roughly) half 
produced in the s-process and half in the r-process.   Nd alone 
does not actually constrain the abundance 
contributions significantly.   

The [Ba/Eu] and [La/Eu] ratios in Sculptor, Fornax and Carina 
clearly increase with metallicity, as in the halo stars.   
This suggests that some level of star formation must have 
continued after any initial, early epoch star burst so that 
subsequently more metal-rich objects could be contaminated 
from the early metal-poor AGB stars.  
Of course, this contamination time scale must be 
{\it greater} than the life time of the AGB stars ($\sim 1$ Gyr).  
The seemingly flat [Ba/Eu] and [La/Eu] ratios in
Leo~I suggest that the contribution from AGB stars
must have been fairly small (from metallicity [Fe/H] 
$= -1.5$ to [Fe/H] $= -1.1$).   Thus the timescales between 
these two epochs should have been fairly short.   A short timescale
between these two epochs would imply that there should be little SN Ia
contribution during this period and thus any decline in the even-Z elements 
would be due
to small star formation events and thus few high mass SN II to produce
alpha elements. 
More stellar abundances in this metallicity range would help to
confirm this suggestion since with only two stars 
cannot rule out a small slope in the [Ba/Eu] and [La/Eu] ratios.

The metal-rich star in Fornax, Fnx-21, shows remarkable
enrichment in all s-process elements (and possibly Eu), 
often greater than the enrichments in the Galactic halo stars 
and clearly shows a super-solar s/r ratio.    
The most likely possibility is that this star underwent 
mass transfer in a binary system with an evolved AGB star.
However, with such a small sample of stars we cannot rule out the 
possibility that the most metal-rich stars in Fornax have had a very large
s-process enrichment from AGB stars in comparison to 
the total number of r-process SN events.  
However, this second hypothesis seems to contradict
the slightly enhanced alpha/Fe ratio and an increasing 
[Eu/H] abundance with increasing metallicity which imply continued 
contribution from SN II.    
Only analyses of additional metal-rich stars in Fornax 
will be able to distinguish between these two possibilities.

\subsection{Carina's Abundance Pattern}

Carina has a bursting star formation history as determined
from its CMD (Hernandez \etal 2000, 
Hurley-Keller, Mateo \& Nemec 1998, 
Smecker-Hane \etal 1994, Mighell 1990).
One might expect to see this signature
in the alpha-element/Fe ratios.   After EACH burst, the
alpha-element ratio increases rapidly (because of the rapid 
influx of alpha rich SN II material; e.g. Gilmore \& Wyse 1991), 
followed by a slow decline as Fe is produced by the SN Ia supernovae.   
Of course, to sample this pattern would require some low level 
of star formation between each burst.   Also, this assumes a
fully sampled IMF, which may not occur (statistically) 
in very low mass star formation events. 

In Carina, we may see this pattern for the first time 
in any galactic system\footnote{Very recent analyses of red 
giants in LMC clusters may also show the alpha/Fe ratios
predicted from bursting star formation history models,
Hill (2002).}.
Our most metal-poor star (Car-10, [Fe/H]=-1.94) 
has slightly enhanced [$\alpha$/Fe] ratio, but the next most metal-poor 
star (Car-3, [Fe/H]=-1.65) has quite a low [$\alpha$/Fe].   
In addition the [$\alpha$/H] ratio
of Car-10 and Car-3 are nearly the same which implies that the increase
in iron peak elements was not accompanied by any detectable increase in 
the alpha elements, as is expected from SN Ia.
The remaining three
objects return to high [$\alpha$/Fe] (perhaps even higher
than Car-10) as predicted if a
burst of star formation followed that polluted the
interstellar medium with alpha-elements.   If this 
interpretation is correct the burst must have happened 
between [Fe/H]=-1.65 and [Fe/H]=-1.60, according to our 
iron abundances;  this is the metallicity range predicted 
$\sim$7 Gyr ago when a burst lasting 2-4 Gyr is predicted
by Hurley-Keller \etal (1998) and Hernadez \etal (2000). 
Further discussions of ages and burst populations are discussed
in Paper II).
 
The pattern repeats in all of the alpha-elements, strongly 
suggesting that this is not a problem with a particular
set of absorption lines (e.g., that Car-3 and its analysis 
is NOT unusual).
We also strongly suggest that this is not a pattern brought 
on by atmospheric parameter uncertainties since the Fe abundances
and temperatures are quite typical.   The only distinction
is that Car-3 has a very low gravity determination, however 
most of the alpha-element ratios are NOT sensitive to gravity 
(e.g., Mg, Si, Ca, Ti, see Table 6) and still show this pattern.

In addition to the alpha-abundance pattern in the five
Carina stars, we find supportive evidence that their
chemical abundances are related to the star formation
history in the s\&r-process ratios as well.   The Ba, La, Nd, and Eu
abundances are all similar to primarily r-process in
the most metal-poor star (Car-10, [Fe/H]=$-$1.94).
In the next star (Car-3, [Fe/H]=$-$1.65), the Ba/Eu, La/Eu,
and Nd/Eu appear to be very slightly larger, suggestive
of some small s-process enrichment;
the SFH by Hurley-Keller \etal 1998 suggests a 3 Gyr
hiatus between the first and second bursts of star
formation, which is sufficient time for AGB stars to
contribute some s-process fraction.   
The next two stars, with [Fe/H]=$-$1.6, show a significant 
increase in La/Eu, and also slight increases in Ba/Eu and Nd/Eu.  
This suggests further AGB contributions.
The increase in their alpha elements implies SN II contributions, 
which could also provide the r-process elements, and
drive down the ratio of s-process to r-process abundances,
but we do not see this.   Possibly very little r-process
elements were formed or were incorporated into the ISM when
these stars formed, or the AGB contributions were simply 
more significant. 
It is important to point out that we are not predicting that 
Car-3 was formed in the second burst.   Car-3 could have been 
formed in an intermediate star formation event some considerable 
time before the second burst and thus further AGB contamination 
could have occurred.   

One difficulty in the interpretation of the chemical evolution 
of Carina as a burst pattern is the flat (or even
declining) [Eu/H] abundances, see Figure 12.   
If we expect significant SN II  between 
the Car-3 and the [Fe/H]$=-1.6$ Carina stars then we might expect 
a rise in the [Eu/H] abundance since Eu (as a primarily r-process element)
is thought to be produced in SN II.    
One possible
explanation for this contradiction is that the mass range of SN that produce
most of the Eu at this metallicity is narrow enough that a low mass IMF
would restrict the number of these events.   Another possibility is that
some of the subsequent r-process material has been lost from the galaxy
(blow-out?).   Since we detect a factor of two increase in the even-Z abundance
the blow-out would have to be very selective.
But a final possibility is simply that we have underestimated our errors 
for the Eu abundance based on this single very weak line and cannot 
detect a subtle increase in Eu that may be present.

Figure 14 is an alternative way to view the entire chemical evolutionary 
history for Carina.  Figure 14
shows the abundance pattern for Car-10, Car-12 and Car-3.
The abundance pattern has been normalized to Mg for the light
elements and normalized to Eu for the heavy elements.    The top panel
shows the solar system abundance pattern as a solid line and the dotted
line represents the abundance pattern implied by SN IIL SN from
Qian and Wasserburg (2002)  while the dashed line represents the 
the r-process abundance pattern from Arlandini \etal (1999).   The solid
line and the dotted line deviate furthest apart in the iron peak, elements
Cr - Ni.    We show the abundance pattern of three Carina stars, filled
squares representing Car 10
(our most metal-poor Carina giant), crosses representing Car 3 
(the Carina giant with the extremely
low alpha to iron abundance ratio), and open squares representing
Car 12 (our most metal-rich Carina giant).    Among the iron peak elements
the Car 3 abundances stand out as anomalous, while the Car 12 and Car 10 
abundances lay between the solar and SN IIL SN abundance patterns.   
Among the heavy elements there appears to be a spread in the abundance pattern
with Car 10 fitting the Arlandini \etal (1999) pure r-process abundance pattern
and Car 12 being closest to a solar abundance pattern.    
Because of the large
dynamic range in the top panel of Figure 14 a comparison between the different
abundance patterns is difficult.

The bottom
panel for Figure 14 shows the same abundance pattern but with the average
globular cluster abundance removed.   Since we only have a single 
globular cluster star without the deep mixing abundance pattern we have
adopted the Mg and O abundances from that star (M55-76) and have excluded
Na and Al.    The
points in the bottom panel of Figure 14 are the same as those given in the
top with the addition of open circles which represent the solar abundance
pattern.  
The elements below atomic number 20, ie. O, Mg, Si are similar to that of the 
globular cluster but the iron peak elements, ie. atomic number 21-30,
are clearly overabundant.   
The X's (Car-3) exhibit the highest overabundance in the iron peak.   
As mentioned before we interpret this
to be a due to a long period of SN Ia contamination before a 
later burst which brings the peak back down (or the alpha elements
back up).     
The fact that the most metal-poor Carina star (Car-10) shows an overabundance
of iron peak elements does not necessarily mean that SN Ia have 
contributed to its abundance pattern since as we have mentioned previously
a low mass star formation event can produce a low alpha,
with respect to the iron peak, abundance pattern.
The heavy elements also show a clear evolution toward
the solar abundance distribution (the open symbol Car-12 is the most
metal-rich in the Car sample and shows the most solar like heavy element
abundance distribution).   We interpret this to mean that significant
time has passed between the formation of each of these dSph stars, 
i.e., to allow subsequent AGB contamination.    

If this burst-like abundance pattern can be supported with other
stars in Carina in this metallicity range (near [Fe/H]=-1.6), 
then this would be the first proof of the theoretical 
bursting galaxy chemical enrichment models.

\section{Discussion} 

The underabundance of the alpha elements (with respect to globular 
cluster stars) found at [Fe/H]$ = -1.5$ can be interpreted in two 
ways;  either as the onset of SN Ia at lower metallcities 
than is found in the halo, or as a small star formation event
where there are very few massive stars (the ones that produce the alpha 
elements).   Since the IMF is similar in nearly every
environment in which it is studied (e.g., Magellanic Clouds and
Galactic clusters, Massey 2003),
then usually the alpha-element ratios is interpreted in terms of
the onset of SN Ia, but the effect of the absence of many massive stars 
in a small star formation event should not be ignored.    
However, for Fornax and Carina where a large spread in ages 
is expected (see Mateo 1998, Hernandez \etal 2000,
Hurley-Keller, Mateo \& Nemec 1998,
Smecker-Hane \etal 1994, Mighell 1990, and Paper II),
then SN Ia contamination should be expected at higher metallicities.

We also note that, if the iron-peak enhancements (as seen in Figure 14) 
are due to SN Ia, over the metallicity range $-2 <$ [Fe/H]$ < -1$, 
and yet the [Mn/Fe] and [Cu/Fe] remain flat, then SN Ia can not be the 
cause of the upturn in Mn and Cu seen among the Galactic halo stars.    
This is also supported by the very low [Cu/alpha] ratios shown
in Figure 10.    As discussed in Section 8.4,
 we suggest that a metallicity-dependent SN yield
 (e.g., SN II, Timmes \etal 1995),
 may be the formation site for Cu and Mn in metal-poor stars.

A similar type of argument can also be made for the source of the 
first s-process peak in metal-poor stars.    Since the timescales
for SN Ia and AGB contamination are similar, and the slopes 
of [Y/H] vs [Fe/H] are different between the Galactic halo stars 
and the dSph stars, then the source for Y in metal-poor stars is 
not SN Ia nor AGB stars.  {\it It must come from another source,
such as SN II (again, possibly a metallicity-dependence).}
The large [Ba/Y] ratio seen in the dSph stars with [Fe/H] $> -1.6$ 
(see Figure 13) might be due to Ba (but not Y) being enhanced by 
the s-process.
The fact that the most metal-rich star in Fornax, Fnx-21, has
Ba/Y that is halo-like is due to increased Y (also seen in
Figure 13), mostly likely because Y has been enhanced by a
greater factor than Ba (since Ba, and La, are also enhanced in
this star) from more metal-rich AGB stars.   If Zn also has a 
small component that is linked to the first s-process elements 
then the slight underabundance of Zn might be linked to the 
underabundance of Y.

\subsection{DSph Abundances and the Galactic Halo}

Several lines of evidence suggest that the Galactic halo is, 
at least partially, composed of accreted dSph galaxies.   These 
include the current assimilation of the Sgr dwarf (Ibata \etal 1997,
Dohm-Palmer \etal 2000, Newberg \etal 2002), and possibly 
Omega Cen (Majewski \etal 2000, assuming that it is a stripped dSph).   
The abundances presented here for the metal-poor stars in four dSph's show a
strong iron peak signature (regardless of the origin) or viewed differently
as low alpha to iron ratio with respect to Galactic halo stars.  
Since the halo's metallicity distribution
peaks near [Fe/H] $= -1.8$ and those stars show a higher alpha to iron ratio
than the dSph stars (see the Figures 3-8 in this work and Figure 12 in 
Fulbright 2002), clearly a large percentage of the 
halo can not have be produced from dSph similar to those 
analyzed here or we would see a many stars with a strong iron peak 
abundance pattern in the halo.    
Fulbright (2002) found that less than 10\% of the 
local metal-poor ([Fe/H] $< -1.2$) stars sample have 
alpha to iron abundance ratios
similar to those found in 
the dSph sampled in this work and SCS01.    However, by subdividing his
sample by total space velocity, the highest space velocity stars have 
systematically lower alpha to iron abundance ratios.  
Stephens' (1999) sample was kinematically
selected to probe the outer halo by looking for high velocity local
stars.   This sample also exhibits
low [Na/Fe] ratios and low even-Z to iron ratios (with respect to the other
halo samples).     At the same metallicities as the Stephens (1999) sample,
our dSph samples have low [Na/Fe] and even lower low even-Z to iron ratios.
Perhaps the disrupted dSph were similar to those studied in this work
contribute to the the high space velocity tail Galactic halo.

Nissen \& Schuster (1997) conducted a detailed abundance analysis of a 
nearby sample of disk and halo stars with similar metallicities to study
the disk-halo transition.   Their sample was chosen to get an equal number
of disk and halo stars as defined by the stars stellar rotation.  Of their
13 chosen halo stars, 8 show an unusual abundance pattern:   low alpha
element to iron ratio, low [Ni/Fe] abundances and low [Na/Fe] abundances.
These odd halo stars also exhibited larger R$_{max}$ and z$_{max}$ orbital
parameters than the other halo stars sampled.    Nissen \& Schuster (1997)
suggest that these anomalous stars may have their origins in disrupted 
dSph.    The dSph stars in our sample at a similar metallicity [Fe/H] $= -1.0$
also exhibit sub-solar [Na/Fe] and [Ni/Fe], and low even-Z to iron 
abundances.   This seems to lend support to the idea put forward by 
Nissen \& Schuster (1997) that a large fraction ($> 50\%$)
of the metal-rich halo may have their origin in disrupted dSph like those
studied in this work.   

This still leaves the question of the origin of the metal-poor 
halo though, and the fraction of the metal-poor halo that formed
through monolithic collapse versus accretion of dSph galaxies.   
We note that we have examined the [alpha/Fe] ratios in a subset of
the dSph stars, that is those with the oldest ages ($\sim$15 Gyr,
from Paper II).  On average, [alpha/Fe] $\sim$ +0.15 with a range
from solar to +0.4.   This average is still lower than the metal-poor
(presumably old) halo stars, yet the range does overlap.   It is likely
that some fraction of the old, metal-poor halo is composed of disrupted
dSphs like those examined here, but we continue to agree with SCS01
that the dSphs cannot account for the majority.

\subsection{Connection to other dSph Galaxies}

There are not a large number of publications with high resolution
detailed abundance analyses of dSph stars.   Bonifacio \etal (2000)
and Smecker-Hane \& McMilliam (2002) have samples of stars in the 
Sagittarius dSph.   Shetrone et al. (1998) analyzed 4 giants in the 
Draco dSph and these results were incorporated into SCS to yield a
sample of 6 giants in Draco, 6 giants in Ursa Minor, and 5 giants in 
Sextans.      

The SCS sample should be the most straight forward to compare with
this work since many of the methods are the same.    The population
sampled in Draco, Ursa Minor
and Sextans contains more very metal-poor ([Fe/H] $< -2$) so 
we shall restrict ourselves to comparisons between $-2 <$ [Fe/H] $< -1$.
The overall abundance distribution differences could be better
addressed in a low resolution abundance population paper.   In this 
restricted metallicity range the Draco, Ursa Minor and Sextans samples
have very similar abundance patterns to the dSph abundance patterns
of Sculptor, Fornax, Leo I and Carina.   This includes under abundant
alpha to iron abundance ratios with respect to the halo, a slightly lower
[Zn/Fe] than found in the halo, a low [Y/Fe] at the slightly higher
metallicities.    The one exception to the similarities is the evolution
from low s-process to r-process ratios to high s-process to r-process ratios
seen in Fornax, Carina and Sculptor and not in Leo, Draco, and Ursa Minor
(unfortunately no Eu abundances were determined by SCS for Sextans).
This single difference is likely due to a star formation history which
does not seem to be linked in any obvious fashion to galaxy mass since
Fornax has the largest mass out of this sample and Carina and Sculptor 
are some of the least massive.    Despite this lingering question it
is comforting that all of these dSph have very similar intermediate
chemical evolutionary histories.

Combining the Sagittarius dSph samples into a single picture
(Bonifacio \etal 2000, Smecker-Hane \& McMilliam 2002) 
reveals a galaxy that seems to be intermediate between the Galactic
halo and the dSphs in this paper.   The metal-poor stars 
([Fe/H] $\sim -1.5$) in this paper and SCS exhibit slightly 
enhanced [alpha/Fe] 
(defined as the average of [Si/Fe], [Ca/Fe] and [Ti/Fe])   
but less than the ratios seen in the Galactic halo.  
For the metal-poor stars in the Sagittarius dSph, [alpha/Fe] 
are slightly higher.
But, as mentioned earlier, the [Ca/Fe] and particularly [Ti/Fe] 
abundances may not be good indicators of the relative contribution 
of SN II to SN Ia since some models of both types of SN produce 
both Si, Ca and Ti in reasonably similar amounts 
(see Woosley \& Weaver 1995 and Table 3 in Iwamoto \etal 1999).  
 
It should be noted that one of the three Smecker-Hane \& McWilliam 
metal-poor stars exhibits a deep mixing abundance pattern.    
However, no metal-poor dSph stars in SCS or
this work show a deep mixing abundance pattern.      

The metal-rich stars in the Sagittarius dSph ([Fe/H] $\sim -0.5$) 
exhibit solar like [alpha/Fe] and slightly enhanced
s-process to r-process ratios of heavy elements.   These metal-rich 
stars also exhibit a large deficiency of Al, Na, Ni and Y. 
Again, the metal-rich stars share some similarities to the 
Nissen \& Shuster (1997) anomalous stars.    
There is little overlap between the metal-rich stars
in the Sagittarius dSph and the other published dSph abundances though; 
only the one star in, Fnx-21, is as metal-rich, but it may be an anomalous 
s-process rich mass transfer star (see above).   Comparisons between
the Sagittarius dSph and the other dSph will have to wait until larger
surveys of the metal-poor Sagittarius dSph and the metal-rich other
dSph are conducted.

\section{Summary }

Certain abundance patterns appear to be very similar between
the four dwarf spheroidal galaxies studied here (the Sculptor,
Fornax, Leo I, and Carina dwarf spheriodals) and the others
examined in the literature (the Ursa Minor, Draco, Sextants, and
Sagittarius dwarf spheriodals).   These include;

\noindent 1.  Galactic halo-like abundances for the iron-group elements,
in particular [Sc/Fe], [Cr/Fe], [Co/Fe], and [Ni/Fe].   In addition,
[Mn/Fe] is halo-like in all the dSph stars.

\noindent 2.  The most metal-poor dSph stars, with [Fe/H] $< -1$, show
halo-like s\&r-process abundance patterns and [Cu/Fe] abundances.    
The only exception is the first peak s-process element, Y, 
where [Y/Fe] is lower than in the halo.

\noindent 3.  The most metal-poor dSph stars, with [Fe/H] $< -1$, show
lower [Zn/Fe] abundance ratios than the Galactic halo stars.

\noindent 4.  None of the stars in the dSphs show the deep mixing
abundance pattern (a possible exception may be one star in Sagittarius).
For example, all of the dSph stars with [Fe/H] $< -1$ show a very low
Na abundance, with [Na/Fe] $\sim -0.4$.

The alpha-element abundance patterns are not similar between the
dSphs though.   The [alpha/Fe] ratio can vary from galaxy to galaxy
and can vary with metallicity in an individual galaxy. 
Specifically, Carina shows a wide dispersion in the [alpha/Fe] ratios
at a given metallicity, which we interpret in terms of its
bursting star formation history.    Sculptor and Leo I show a
slightly declining alpha abundance pattern with increasing metallicity,
as do Sextants, Ursa Minor, and Sagittarius.
Fornax and Draco show a roughly constant alpha abundance
over the metallicites sampled.
The alpha/Fe ratios in the dSph stars continue to be lower than
seen in Galactic halo stars of similar metallicity, thus we remain
in agreement with Shetrone \etal (2001) that the majority of the
Galactic halo cannot have formed from disrupted dSph systems.
However, similarities in the [Ni/Fe] and [Na/Fe] abundances with
high velocity halo stars from Nissen \& Schuster (1997) may suggest
that as much as 50\% of the metal-{\it rich} halo is comprised of
dSph stars.

Despite the generally halo-like s\&r-process abundances in the
metal-poor stars (above), not every dSph exhibits the same evolution
in the s\&r-process abundance pattern.  Carina, Sculptor and Fornax
show a rise in the s/r-process ratio with increasing metallicity,
evolving from a pure r-process ratio to a solar-like s\&r-process ratio.
On the other hand, Leo I, Draco, and Ursa Minor appears to show an
r-process dominated ratio over the range in metallicities sampled.
Again, we attribute this to differences in the star formation
histories of these galaxies.

The dSph abundances place new constraints on nucleosynthetic
origins of several elements.   We find that [Cu/Fe] and [Cu/alpha]
are flat over a large range in metallicity in all of the dSph stars.
We take these abundance ratios in combination with the known age
spread in several of the dSphs as evidence for a metallicity
dependent SN (Ia or II) yield for Cu.  The same is found for Mn.
Also, we attribute differences in the evolution of [Y/Fe] in the dSph
stars versus the halo stars to a very weak AGB or SN Ia yield of Y
(especially compared to Ba).    That a lower and flatter Ba/Y ratio
is seen in the halo is due to the pattern being erased by the large
metallicity dispersion in the halo (as described by McWilliam 1997).
If Zn also has a small component that is linked to the production of
the first s-process elements, then the slight underabundance of Zn
might be linked to the underabundances in Y.

\acknowledgments

We thank the Paranal Observatory staff for excellent support we received
during both of our visitor runs.  ET gratefully acknowledges support from
a fellowship of the Royal Netherlands Academy of Arts and Sciences, and 
PATT travel support from University of Oxford.    KAV would like to 
thank the NSF for support through a CAREER award, AST-9984073.   
We thank Sonya M. Clarkson, Christina M. Blank, Fitih M. Mohammed,
and Leah E. Simon for their assistance with the line measurements and
the atmospheric analyses.
We thank A. McWilliam for useful discussions and an invaluable review article.

\clearpage
%\epsscale{0.8}
%\plotone{shetrone.fig1.eps}
\figcaption{A comparison the EW from this work and Minniti et al. 1993.
The triangles represent M30 D lines, the squares represent M55 283 lines,
the crosses represent M55 76 and the circles represent M68 53 lines.  The
solid line is the 45 degree line. The dotted line is offset from the 45
degree line by an error of 6 m\AA.    The dashed line represents a 10\% error
convolved with the 6m\AA error.
\label{fig1}}

%\clearpage
%\epsscale{0.8}
%\plotone{shetrone.fig2.eps}
\figcaption{Na and Al abundances for our sample:
Carina (red squares), the Sculptor (blue circles), Fornax (green triangles)
Leo I (magenta pentagons), and the globular cluster abundances (large
open squares).  
The small symbols are taken
from the literature to represent the disk, and halo populations:
Edvardsson \etal 1993 (small circles), Nissen \& Schuster 1997 (small stars),
Stephens 1999 (small pentagons), Gratton \& Sneden 1988 (small squares),
and McWilliam 1995 (small triangles).   
The errorbars presented here are
the systematic errors 
in Tables 8-11 and the internal errors from Table 7 added in quadrature.
\label{fig2}}

%\clearpage
%\epsscale{0.8}
%\plotone{shetrone.fig3.eps}
\figcaption{Carina [O/Fe], [Mg/Fe] and [Si/Fe] abundances (red squares)
are plotted against
metallicity.  The symbol types are the same as Figure 2 with the addition
of Gratton \& Sneden 1991 and Gratton \& Sneden 1994 (small squares).
The errorbars presented here are
the systematic errors 
in Tables 8-11 and the internal errors from Table 7 added in quadrature.
\label{fig3}}

%\clearpage
%\epsscale{0.8}
%\plotone{shetrone.fig4.eps}
\figcaption{Sculptor [O/Fe], [Mg/Fe] and [Si/Fe] abundances 
(blue circles) are plotted against
metallicity.  The symbol types are the same as Figure 3.
The errorbars presented here are
the systematic errors 
in Tables 8-11 and the internal errors from Table 7 added in quadrature.
\label{fig4}}

%\clearpage
%\epsscale{0.8}
%\plotone{shetrone.fig5.eps}
\figcaption{Fornax (green triangles) and Leo (magenta pentagons) 
[O/Fe], [Mg/Fe] and [Si/Fe] abundances are plotted against
metallicity.  The symbol types are the same as Figure 3.
The errorbars presented here are
the systematic errors 
in Tables 8-11 and the internal errors from Table 7 added in quadrature.
\label{fig5}}

%\clearpage
%\epsscale{0.8}
%\plotone{shetrone.fig6.eps}
\figcaption{Carina (red squares) [Ca/Fe], [TiI/Fe] and [TiII/Fe] 
abundances are plotted against
metallicity.  The symbol types are the same as Figure 3.
The errorbars presented here are
the systematic errors 
in Tables 8-11 and the internal errors from Table 7 added in quadrature.
\label{fig6}}

%\clearpage
%\epsscale{0.8}
%\plotone{shetrone.fig7.eps}
\figcaption{Sculptor (blue circles) 
[Ca/Fe], [TiI/Fe] and [TiII/Fe] abundances are plotted against
metallicity.  The symbol types are the same as Figure 3.
The errorbars presented here are
the systematic errors 
in Tables 8-11 and the internal errors from Table 7 added in quadrature.
\label{fig7}}

%\clearpage
%\epsscale{0.8}
%\plotone{shetrone.fig8.eps}
\figcaption{Fornax (green triangles) and Leo (magenta pentagons) 
[Ca/Fe], [TiI/Fe] and [TiII/Fe] abundances are plotted against
metallicity.  The symbol types are the same as Figure 3.
The errorbars presented here are
the systematic errors 
in Tables 8-11 and the internal errors from Table 7 added in quadrature.
\label{fig8}}

%\clearpage
%\epsscale{0.8}
%\plotone{shetrone.fig9.eps}
\figcaption{Iron peak abundances for our sample:
Carina (red squares), the Sculptor (blue circles), 
Fornax (green triangles)
Leo I (magenta pentagons), and the globular cluster abundances (large
open squares).
The symbol small types are the same as Figure 3 with the addition of
Sneden \etal 1991 (small squares) and Primas \etal 2000 (crosses).
The errorbars presented here are
the systematic errors 
in Tables 8-11 and the internal errors from Table 7 added in quadrature.
\label{fig9}}

%\clearpage
%\epsscale{0.8}
%\plotone{shetrone.fig10.eps}
\figcaption{  Mn and Cu abundances for our sample:
Carina (red squares), the Sculptor (blue circles), Fornax (green triangles)
Leo I (magenta pentagons), and the globular cluster abundances (large
open squares).  The small symbols are taken
from the literature to represent the disk, and halo populations:
Gratton 1989 (small squares), McWilliam 1995 (small triangles)
Gratton \& Sneden 1988 (small squares), Sneden \etal 1991 (small squares)
and Primas \etal 2000 (crosses).   $\alpha$ is defined as the average of the
Mg and Ca abundances.
The errorbars presented here are
the systematic errors 
in Tables 8-11 and the internal errors from Table 7 added in quadrature.
\label{fig10}}

%\clearpage
%\epsscale{0.8}
%\plotone{shetrone.fig11.eps}
\figcaption{The Y and Ba abundances for our sample:
Carina (red squares), the Sculptor (blue circles), Fornax (green triangles)
Leo I (magenta pentagons), and the globular cluster abundances (large
open squares).
The symbol small types are the same as Figure 10 with the 
substitution of Burris \etal 2000 (small crosses).   
The errorbars presented here are
the systematic errors 
in Tables 8-11 and the internal errors from Table 7 added in quadrature.
\label{fig11}}

%\clearpage
%\epsscale{0.8}
%\plotone{shetrone.fig12.eps}
\figcaption{The Eu abundances for our sample:
Carina (red squares), the Sculptor (blue circles), Fornax (green triangles)
Leo I (magenta pentagons), and the globular cluster abundances (large
open squares).  
The small symbol types are the same as Figure 9 with the addition
of Eu for the Edvardsson \etal 1993 coming from Koch \& Edvardsson 2002 and
the small crosses representing data from Burris \etal 2000.   
The errorbars presented here are
the systematic errors 
in Tables 8-11 and the internal errors from Table 7 added in quadrature.
\label{fig12}}

%\clearpage
%\epsscale{0.8}
%\plotone{shetrone.fig13.eps}
\figcaption{The s\&r-process element ratios for our sample:
Carina (red squares), the Sculptor (blue circles), Fornax (green triangles)  
Leo I (magenta pentagons), and the globular cluster abundances (large  
open squares). 
The small symbol types are the same as Figure 10 with the addition
of Eu for the Edvardsson \etal 1993 coming from Koch \& Edvardsson 2002
and the small crosses representing data from Burris \etal 2000.  
The dotted line
represents the pure r-process abundance ratios from Burris \etal 2000.   The
solid line represents the pure r-process abundance ratios from 
Arlandini \etal 1999.
The errorbars presented here are
the systematic errors 
in Tables 8-11 and the internal errors from Table 7 added in quadrature.
\label{fig13}}

%\clearpage
%\epsscale{0.7}
%\plotone{shetrone.fig14.eps}
\figcaption{The top panel shows the abundance patterns for Car 10 
(red filled squares), Car 12 (red open squares) and Car3 (red crosses) 
normalized
to the Mg abundance, for the light elements and Eu for the heavy elements.   
The solid line is solar abundance pattern, the
dotted line is the predicted SN IIL abundance pattern from 
Qian \& Wasserburg (2002)  and the dashed line is the pure r-process 
abundance pattern from Arlandini \etal (1999).    The bottom panel 
shows the residual abundances pattern for the same three stars after subtracting
off our observed globular cluster abundance pattern.  The open circles 
represent the solar system abundances.   The Na and Al abundances are
excluded in the bottom panel.
The errorbars presented here are
the systematic errors 
in Tables 8-11 and the internal errors from Table 7 added in quadrature.
\label{fig14}}

\clearpage
\begin{deluxetable}{llllccc}
\tablenum{1}
\footnotesize
\tablecaption{Observations}
\tablewidth{0pt}
\tablehead{
\colhead{Date} & 
\colhead{Begin} &  
\colhead{Object} & 
\colhead{Exp} & 
\colhead{Airmass} & 
\colhead{DIMM\footnotemark[1]} &
\colhead{Comments} \nl
\colhead{} &
\colhead{UT} & 
\colhead{} & 
\colhead{(secs)} & 
\colhead{} & 
\colhead{arcsec} & 
\colhead{} 
} 
\startdata
16Aug2000 & 04:24  & Scl-H461 & 3600 & 1.36 & 0.92 &        \\
16Aug2000 & 05:26  &          & 3600 & 1.15 & 0.98 &        \\
16Aug2000 & 06:29  &          & 3600 & 1.05 & 0.66 &        \\
16Aug2000 & 07:34  & Scl-H400 & 1945 & 1.02 & 0.46 &tracking\\
16Aug2000 & 09:03  &          & 3600 & 1.09 & 0.55 &        \\
17Aug2000 & 04:53  &          & 3600 & 1.23 & 0.58 &        \\
17Aug2000 & 05:56  &          & 3600 & 1.09 & 0.56 &        \\
18Aug2000 & 04:18  & Scl-H479 & 3600 & 1.35 & 0.54 &        \\
\enddata
\footnotetext[1]{ This is just an indication of the external seeing
measured automatically by the seeing monitor (DIMM) on the mountain.
Usually the seeing on the instrument is better than this.}
\footnotetext[2]{DIMM wasn't working for a short period}
\end{deluxetable}

\clearpage

\begin{deluxetable}{lllllll}
\tablenum{2}
\footnotesize
\tablecaption{The Stellar Sample}
\tablewidth{0pt}
\tablehead{
\colhead{Galaxy} & \colhead{star id} &  \colhead{V} & 
\colhead{(B--V)$_o$} & \colhead{(V--I)$_o$} &
\colhead{RV$_{\rm helio}$} & \colhead{ref} \\[.2ex]
\colhead{} & \colhead{} &  \colhead{} & \colhead{} & \colhead{} & 
\colhead{(km/s)} & \colhead{} \\
} 
\startdata
Sculptor    &    H400   &  18.30        &   0.89   & 1.04  &  109   & 1,2,11,12  \\
            &    H459   &  18.14        &   1.03   & 1.16  &  116  &     \\
            &    H461   &  17.56        &   1.17   & 1.35  &  104  &     \\
            &    H479   &  17.23        &   1.20   & 1.31  &  98   &     \\
            &    H482   &  17.65        &   1.21   & 1.41  &  107  &     \\
Fornax      &    M12    &  18.43        &   1.33   & 1.50  &  54   &  3,11,12  \\
            &    M25    &  18.59        &   1.49   & 1.53  &  49   &     \\
            &    M21    &  18.37        &   1.59   & 1.66  &  53   &     \\
\enddata
\tablecomments{ 1. Hodge 1965; 2. Queloz, Dubath, \& Pasquini 1995;
3. Mateo et al. 1991; 4. Mateo et al. 1993; 5. Mateo et al. 1998; 
6. Harris 1996, 7. Harris 1975, 8. Alcaino 1975, 
9. Alcaino \& Liller 1980; 10. Alcaino 1977; 11. Tolstoy \etal in prep;
12. E(V-I) from Dean, Warren \& Cousins 1978 }
\end{deluxetable}

\clearpage

\begin{deluxetable}{lcccc} 
\tablenum{5}
\footnotesize
\tablecaption{Atmospheric Parameters}
\tablewidth{0pt}
\tablehead{
\colhead{OBJ} & \colhead{\teff} & \colhead{\logg} & \colhead{$\xi$} & \colhead{[FeI/H]} \\[.2ex]
\colhead{} & \colhead{ (K) } & \colhead{} & \colhead{(\kms)} & \colhead{} 
} 
\startdata
M30-D   & 4400 & 0.50 & 2.0  & -2.30 \\
M55-283 & 4600 & 1.20 & 1.65 & -1.75 \\
M55-76  & 4550 & 0.90 & 1.9  & -1.99 \\
M68-53  & 4300 & 0.30 & 2.0  & -2.21 \\
\\
Scl-400 & 4650 & 0.90 & 1.7 & -1.98 \\
Scl-459 & 4500 & 1.00 & 1.65 & -1.66 \\
Scl-461 & 4500 & 1.20 & 1.7 & -1.56 \\
Scl-479 & 4325 & 0.70 & 1.7 & -1.77 \\
Scl-482 & 4400 & 1.10 & 1.7 & -1.24 \\
\\
Fnx-25 & 4025 & 0.00 & 2.0 & -1.21 \\
Fnx-12 & 4150 & 0.00 & 2.1 & -1.60 \\
Fnx-21 & 4000 & 0.50 & 1.7 & -0.67 \\
\\
Leo-2 & 4200 & 0.50 & 1.85 & -1.06 \\
Leo-5 & 4250 & 0.80 & 2.2  & -1.52 \\
\\
Car-2 & 4250 & 0.55 & 2.1  & -1.60 \\
Car-3 & 4250 & 0.20 & 2.2  & -1.65 \\
Car-4 & 4200 & 0.40 & 2.1  & -1.59 \\
Car-10 & 4375 & 0.40 & 2.0 & -1.94 \\
Car-12 & 4300 & 0.60 & 1.9 & -1.41 \\ 
\enddata
\end{deluxetable}
 
\clearpage

\begin{deluxetable}{lrrrrrc} 
\tablenum{6}
\footnotesize
\tablecaption{Abundance Uncertainties for Car-2}
\tablewidth{0pt}
\tablehead{
\colhead{Elem} & \colhead{$\Delta$\teff} & \colhead{$\Delta$\logg} & \colhead{$\Delta\xi$} & \colhead{[M/H]} & \colhead{Cont.} & \colhead{$\Delta$\teff,$\Delta$\logg,$\Delta\xi$}\\[.2ex]
\colhead{}     & \colhead{ $-$100\,K }  & \colhead{ $-$0.2} & \colhead{$-$0.2\,\kms} & \colhead{$-$0.15} & \colhead{4m\AA} & 
\colhead{$-$100,$-$0.25,$-$0.1}
} 
\startdata
{[FeI/H]}      & -0.11 & +0.02 & +0.11 & +0.02 & -0.06 & -0.05 \\
{[FeII/H]}     & +0.10 & -0.07 & +0.05 & -0.03 & -0.08 & +0.03 \\
{[OI/FeI]}    & +0.11 & -0.11 & -0.11 & -0.07 & -0.01 & -0.05 \\
{[NaI/FeI]}   & +0.02 & +0.01 & -0.10 & +0.00 & -0.02 & +0.00 \\
{[MgI/FeI]}   & +0.03 & +0.03 & -0.03 & +0.01 & +0.01 & +0.05 \\
{[SiI/FeI]}   & +0.11 & -0.02 & -0.11 & -0.02 & -0.09 & +0.04 \\
{[CaI/FeI]}   & -0.03 & +0.01 & -0.03 & +0.01 & +0.00 & -0.01 \\
{[ScII/FeI]}  & +0.13 & -0.09 & -0.10 & -0.06 & -0.02 & -0.02 \\
\enddata
\tablecomments{The last column is the abundance uncertainty when 
\teff\ is changed by $-$100K {\it and} the corresponding changes
that would occur in \logg\ and microturbulence are taken into 
account holistically. }
\end{deluxetable}
 
\clearpage

\begin{deluxetable}{lrr} 
\tablenum{7}
\footnotesize
\tablecaption{Adopted Internal Abundance Uncertainties}
\tablewidth{0pt}
\tablehead{
\colhead{Elem} & \colhead{$\sigma$dSph} & \colhead{$\sigma$STND}
} 
\startdata
{[OI/FeI]}    & 0.10 & 0.10 \\
{[NaI/FeI]}   & 0.05 & 0.05 \\
{[MgI/FeI]}   & 0.05 & 0.05 \\
{[AlI/FeI]}   & 0.07 & 0.06 \\
{[SiI/FeI]}   & 0.08 & 0.05 \\
{[CaI/FeI]}   & 0.02 & 0.02 \\
{[ScII/FeI]}  & 0.07 & 0.07 \\
{[TiI/FeI]}   & 0.07 & 0.07 \\
{[TiII/FeI]}  & 0.08 & 0.08 \\
{[CrI/FeI]}   & 0.05 & 0.05 \\
{[MnI/FeI]}   & 0.07 & 0.07 \\
{[FeI/H]}     & 0.07 & 0.06 \\
{[FeII/H]}    & 0.11 & 0.08 \\
{[CoI/FeI]}   & 0.06 & 0.06 \\
{[NiI/FeI]}   & 0.05 & 0.05 \\
{[CuI/FeI]}   & 0.09 & 0.09 \\
{[ZnI/FeI]}   & 0.13 & 0.13 \\
{[YII/FeI]}   & 0.06 & 0.06 \\
{[BaII/FeI]}  & 0.08 & 0.08 \\
{[NdII/FeI]}  & 0.06 & 0.06 \\
{[LaII/FeI]}  & 0.14 & 0.16 \\
{[EuII/FeI]}  & 0.09 & 0.08 \\
{[YII/H]}  & 0.15 & 0.10 \\
{[BaII/YII]}  & 0.04 & 0.04 \\
{[EuII/H]}  & 0.05 & 0.05 \\
{[BaII/EuII]}  & 0.10 & 0.09 \\
{[LaII/EuII]}  & 0.04 & 0.04 \\
{[NdII/EuII]}  & 0.08 & 0.07 \\
\enddata
\tablecomments{The average internal errors are derived from a combination
of the continuum uncertainties, metallicity uncertainties, and the
stellar parameter uncertainties.   The globular cluster standard stars
often have smaller internal errors since the uncertainties in the
continuum placement are significantly smaller.  }
\end{deluxetable}

\clearpage

\begin{deluxetable}{cc|l|l|l|l|l} 
\tablenum{8}
\footnotesize
\tablecaption{Globular Cluster Star Abundances }
\tablewidth{0pt}
\tablehead{
\colhead{Elem} & \colhead{SUN\tablenotemark{*}} &
\colhead{} & 
\colhead{M30-D} &
\colhead{M68-53} &
\colhead{M55-76} &
\colhead{M55-283}   \\
\colhead{} & \colhead{} &
\colhead{} & 
\colhead{AVG ($\Delta_{r}$) } & 
\colhead{AVG ($\Delta_{r}$) } & 
\colhead{AVG ($\Delta_{r}$) } & 
\colhead{AVG ($\Delta_{r}$) }  
} 
\startdata
Fe  & 7.52 & [FeI/H]    &$-2.30$ (0.01) 62  &$ -2.21$  (0.02) 69&$ -1.99$  (0.01) 65  &$ -1.75$  (0.02) 71\\
--- & ---  & [FeII/H]   &$-2.32$ (0.04)S    &$ -2.24$  (0.04)S  &$ -1.98$  (0.03)I    &$ -1.77$  (0.04)S  \\
O   & 8.83 & [OI/FeI]   &$+0.26$  (0.11)I    &$ +0.19$  (0.10)I  &$ +0.48$  (0.08)I    &$ -0.22$  (0.13)I  \\
Na  & 6.33 & [NaI/FeI]  &$+0.27$ (0.08)I    &$ +0.21$  (0.10)S  &$ -0.10$  (0.09)S    &$ +0.12$  (0.08)I  \\
Mg  & 7.58 & [MgI/FeI]  &$+0.52$ (0.09)S    &$ +0.50$  (0.11)S  &$ +0.54$  (0.10)S    &$ +0.11$  (0.11)S  \\
Al  & 6.47 & [AlI/FeI]  &$+1.13$ (0.08)E    &\nodata            &$<+0.38$             &$ +1.15$  (0.09)I  \\
Si  & 7.55 & [SiI/FeI]  &$+0.50$ (0.11)I    &$ +0.66$  (0.13)S  &$ +0.49$  (0.11)S    &$ +0.42$  (0.10)S  \\
Ca  & 6.36 & [CaI/FeI]  &$+0.36$ (0.04)S    &$ +0.29$  (0.05)I  &$ +0.38$  (0.05)S    &$ +0.30$  (0.04)I  \\
\enddata
\tablenotetext{*}{Solar abundances are from Grevesse \& Sauval (1998).}
\tablenotetext{}{ The errors quoted here represent only the random errors. 
These are computed by one of three methods (see text): \\
I $=$ This random error assumes that the random error of the lines of this 
species behave in a way similiar to those of Fe I.  $\sigma_{Fe}/\sqrt{N}$\\
S $=$ This random error is based on the standard deviation of the 
abundance of this species. $\sigma_{el}/\sqrt{N}$\\
E $=$ This random error is based upon an error computed from the 
suggested EW error for the these lines.  See Cayrel (1988). \\
For iron, the number of lines used to compute the standard deviation 
of the mean is given. }
\end{deluxetable}
 
\clearpage

\begin{deluxetable}{cc|l|l|l|l|l|l} 
\tablenum{9}
\footnotesize
\tablecaption{Carina Abundances }
\tablewidth{0pt}
\tablehead{
\colhead{Elem} & \colhead{SUN} &
\colhead{} & 
\colhead{Car-4} &
\colhead{Car-3} &
\colhead{Car-2} &
\colhead{Car-12} &
\colhead{Car-10}   \\
\colhead{} & \colhead{} &
\colhead{} & 
\colhead{AVG ($\Delta_{r}$) } & 
\colhead{AVG ($\Delta_{r}$) } & 
\colhead{AVG ($\Delta_{r}$) } & 
\colhead{AVG ($\Delta_{r}$) } & 
\colhead{AVG ($\Delta_{r}$) }  
} 
\startdata
Fe  & 7.52 & [FeI/H]    &$ -1.59$ (0.02) 71  &$ -1.65$  (0.02) 74 &$ -1.60$  (0.02) 74&$ -1.41$  (0.02) 75  &$ -1.94$  (0.02) 66\\
--- & ---  & [FeII/H]   &$ -1.60$ (0.05)S    &$ -1.63$  (0.04)I   &$ -1.61$  (0.04)I  &$ -1.38$  (0.04)S    &$ -1.94$  (0.04))I  \\
O & 8.83 & [OI/FeI]     &$ +0.22$ (0.09)I    &$ +0.04$  (0.12)I   &$ +0.44$  (0.12)S  &$ +0.17$  (0.10)I    &$ +0.08$  (0.19)I  \\
Na  & 6.33 & [NaI/FeI]  &$ -0.35$ (0.08)I    &$ -0.58$  (0.17)I   &$ -0.38$  (0.10)I  &$ -0.26$  (0.11)S    &$ -0.66$  (0.15)I  \\
Mg  & 7.58 & [MgI/FeI]  &$ +0.26$ (0.09)S    &$ -0.27$  (0.12)I   &$ +0.23$  (0.10)I  &$ +0.24$  (0.10)I    &$ +0.06$  (0.11)I  \\
Al  & 6.47 & [AlI/FeI]  &$ +0.20$ (0.13)I    &$<+0.27$            &$<+0.24$           &$ +0.03$  (0.16)E    &\nodata            \\
Si  & 7.55 & [SiI/FeI]  &$ +0.25$ (0.06)S    &$ -0.28$  (0.17)I   &$ +0.18$  (0.07)I  &$ +0.22$  (0.07)S    &$ +0.38$  (0.22)I  \\
Ca  & 6.36 & [CaI/FeI]  &$ +0.14$ (0.04)I    &$ -0.10$  (0.06)I   &$ +0.20$  (0.05)I  &$ +0.12$  (0.05)I    &$ -0.02$  (0.05)I  \\
\enddata
\tablenotetext{}{See Table 8 for Table Notes }
\end{deluxetable}
 
\clearpage

\begin{deluxetable}{cc|l|l|l|l|l|l} 
\tablenum{10}
\footnotesize
\tablecaption{Sculptor Abundances }
\tablewidth{0pt}
\tablehead{
\colhead{Elem} & \colhead{SUN} &
\colhead{} & 
\colhead{Scl-459} &
\colhead{Scl-479} &
\colhead{Scl-461} &
\colhead{Scl-482} &
\colhead{Scl-400}   \\
\colhead{} & \colhead{} &
\colhead{} & 
\colhead{AVG ($\Delta_{r}$) } & 
\colhead{AVG ($\Delta_{r}$) } & 
\colhead{AVG ($\Delta_{r}$) } & 
\colhead{AVG ($\Delta_{r}$) } & 
\colhead{AVG ($\Delta_{r}$) }  
} 
\startdata
Fe  & 7.52 & [FeI/H]    &$ -1.66$ (0.02) 68  &$ -1.77$  (0.02) 67 &$ -1.56$  (0.02) 67&$ -1.24$  (0.02) 67  &$ -1.98$  (0.03) 45\\
--- & ---  & [FeII/H]   &$ -1.65$ (0.04)I    &$ -1.79$  (0.05)S   &$ -1.58$  (0.04)I  &$ -1.26$  (0.05)I    &$ -1.94$  (0.06))I  \\
O   & 8.83 & [OI/FeI]   &$ +0.22$ (0.15)I    &$ +0.48$  (0.11)I   &$ +0.44$  (0.16)I  &$ +0.18$  (0.18)S    &$      $  \nodata  \\
Na  & 6.33 & [NaI/FeI]  &$ -0.33$ (0.14)S    &$ -0.59$  (0.11)I   &$ -0.55$  (0.11)I  &$ -0.55$  (0.13)I    &$ -0.16$  (0.20)I  \\
Mg  & 7.58 & [MgI/FeI]  &$ +0.36$ (0.13)S    &$ +0.26$  (0.16)S   &$ +0.18$  (0.11)I  &$ +0.09$  (0.13)I    &$ +0.37$  (0.12)I  \\
Al  & 6.47 & [AlI/FeI]  &$<+0.30$            &$<+0.30$            &$<+0.19$           &$<-0.02$             &\nodata            \\
Si  & 7.55 & [SiI/FeI]  &$ +0.22$ (0.15)I    &$ +0.00$  (0.22)I   &$ +0.14$  (0.16)I  &$ -0.07$  (0.15)S    &$      $  \nodata  \\
Ca  & 6.36 & [CaI/FeI]  &$ +0.24$ (0.05)I    &$ +0.17$  (0.05)I   &$ +0.22$  (0.06)S  &$ +0.06$  (0.06)I    &$ +0.38$  (0.09)S  \\
\enddata
\tablenotetext{}{See Table 8 for Table Notes }
\end{deluxetable}
 
\clearpage

\begin{deluxetable}{cc|l|l|l|l|l|l} 
\tablenum{11}
\footnotesize
\tablecaption{Fornax and Leo Abundances }
\tablewidth{0pt}
\tablehead{
\colhead{Elem} & \colhead{SUN} &
\colhead{} & 
\colhead{Fnx-12} &
\colhead{Fnx-25} &
\colhead{Fnx-21} &
\colhead{Leo-5} &
\colhead{Leo-2}   \\
\colhead{} & \colhead{} &
\colhead{} & 
\colhead{AVG ($\Delta_{r}$) } & 
\colhead{AVG ($\Delta_{r}$) } & 
\colhead{AVG ($\Delta_{r}$) } & 
\colhead{AVG ($\Delta_{r}$) } & 
\colhead{AVG ($\Delta_{r}$) }  
} 
\startdata
Fe  & 7.52 & [FeI/H]    &$ -1.60$ (0.02) 48  &$ -1.21$  (0.02) 64 &$ -0.67$  (0.03) 55&$ -1.52$  (0.02) 67  &$ -1.06$  (0.02) 55\\
--- & ---  & [FeII/H]   &$ -1.59$ (0.05)S    &$ -1.17$  (0.04)I   &$ -0.73$  (0.13)S  &$ -1.48$  (0.05)I    &$ -1.10$  (0.05))I  \\
O   & 8.83 & [OI/FeI]      &$ -0.02$ (0.18)S    &$ +0.17$  (0.11)I   &$ +0.12$  (0.17)I  &$ +0.53$  (0.13)I    &$ -0.04$  (0.13)I  \\
Na  & 6.33 & [NaI/FeI]  &$ -0.51$ (0.08)I    &$ -0.31$  (0.08)I   &$ +0.02$  (0.12)I  &$ -0.43$  (0.13)I    &$ -0.36$  (0.10)I  \\
Mg  & 7.58 & [MgI/FeI]  &$ +0.09$ (0.07)I    &$ +0.02$  (0.09)I   &$ +0.20$  (0.12)I  &$ +0.12$  (0.13)I    &$ -0.19$  (0.13)I  \\
Al  & 6.47 & [AlI/FeI]  &\nodata             &$ +0.09$  (0.16)I   &$ -0.04$  (0.24)I  &$ +0.42$  (0.18)I    &$ -0.28$  (0.18)I  \\
Si  & 7.55 & [SiI/FeI]  &$ +0.29$ (0.11)S    &$      $  \nodata   &$ +0.08$  (0.11)I  &$      $  \nodata    &$ +0.00$  (0.13)I  \\
Ca  & 6.36 & [CaI/FeI]  &$ +0.23$ (0.06)I    &$ +0.21$  (0.06)I   &$ +0.23$  (0.08)I  &$ +0.15$  (0.06)I    &$ +0.02$  (0.06)I  \\
\enddata
\tablenotetext{}{See Table 8 for Table Notes }
\end{deluxetable}
 
\clearpage

\begin{deluxetable}{lccc} 
\tablenum{12}
\footnotesize
\tablecaption{Comparisons between MOOG/MARCS and ATLAS/WIDTH for Scl-459}
\tablewidth{0pt}
\tablehead{
\colhead{Elem} & \colhead{MARCS/MOOG} & \colhead{ATLAS/WIDTH} & \colhead{ATLAS/WIDTH}
\\[.2ex]
\colhead{} & \colhead{ (This paper) } & \colhead{ } & \colhead{\& VALD gf's}
} 
\startdata
O I    &  7.39        &    7.32        &    7.39         \\
Na I   &  4.34 (0.20) &    4.42 (0.14) &    4.46 (0.10)  \\
Mg I   &  6.28 (0.20) &    6.26 (0.08) &    6.31 (0.02)  \\
Al I   & $<$5.11      &    $<$5.19     &    $<$4.96      \\
Si I   &  6.11        &    6.18        &    6.18         \\
Ca I   &  4.94 (0.15) &    4.94 (0.10) &    4.99 (0.11)  \\
Sc II  &  1.45        &    1.42        &    1.53         \\
Ti I   &  3.28 (0.12) &    3.18 (0.27) &    3.27 (0.23)  \\
Ti II  &  3.32 (0.15) &    3.23 (0.15) &    3.40 (0.05)  \\
V I    &  2.40 (0.51) &    2.48 (0.35) &    2.48 (0.33)  \\
Cr I   &  3.80 (0.04) &    3.78 (0.02) &    3.78 (0.02)  \\
Mn I   &  3.39 (0.08) &    3.57 (0.08) &    3.57 (0.08)  \\
Fe I   &  5.86 (0.15) &    5.98 (0.15) &    5.97 (0.15)  \\
Fe II  &  5.87 (0.15) &    5.78 (0.18) &    5.74 (0.15)  \\
Co I   &  3.39        &    3.49        &    3.49         \\
Ni I   &  4.70 (0.18) &    4.75 (0.19) &    4.78 (0.23)  \\
Cu I   &  1.50 (0.15) &    2.03 (0.09) &    2.03 (0.08)  \\ 
Zn I   &  3.11        &    3.05        &    3.02         \\
Y II   &  0.53 (0.21) &    0.47 (0.16) &    0.47 (0.16)  \\
Ba II  &  0.80 (0.15) &    0.75 (0.13) &    0.74 (0.13)  \\
La II  & $-$0.52 (0.09) & $-$0.52 (0.06) & $-$0.52 (0.03)  \\  
Nd II  &  0.19 (0.11) &    0.16 (0.07) &    0.20 (0.01)  \\
Eu II  & $-$0.52      & $-$0.54        & $-$0.54         \\
\enddata
\end{deluxetable}

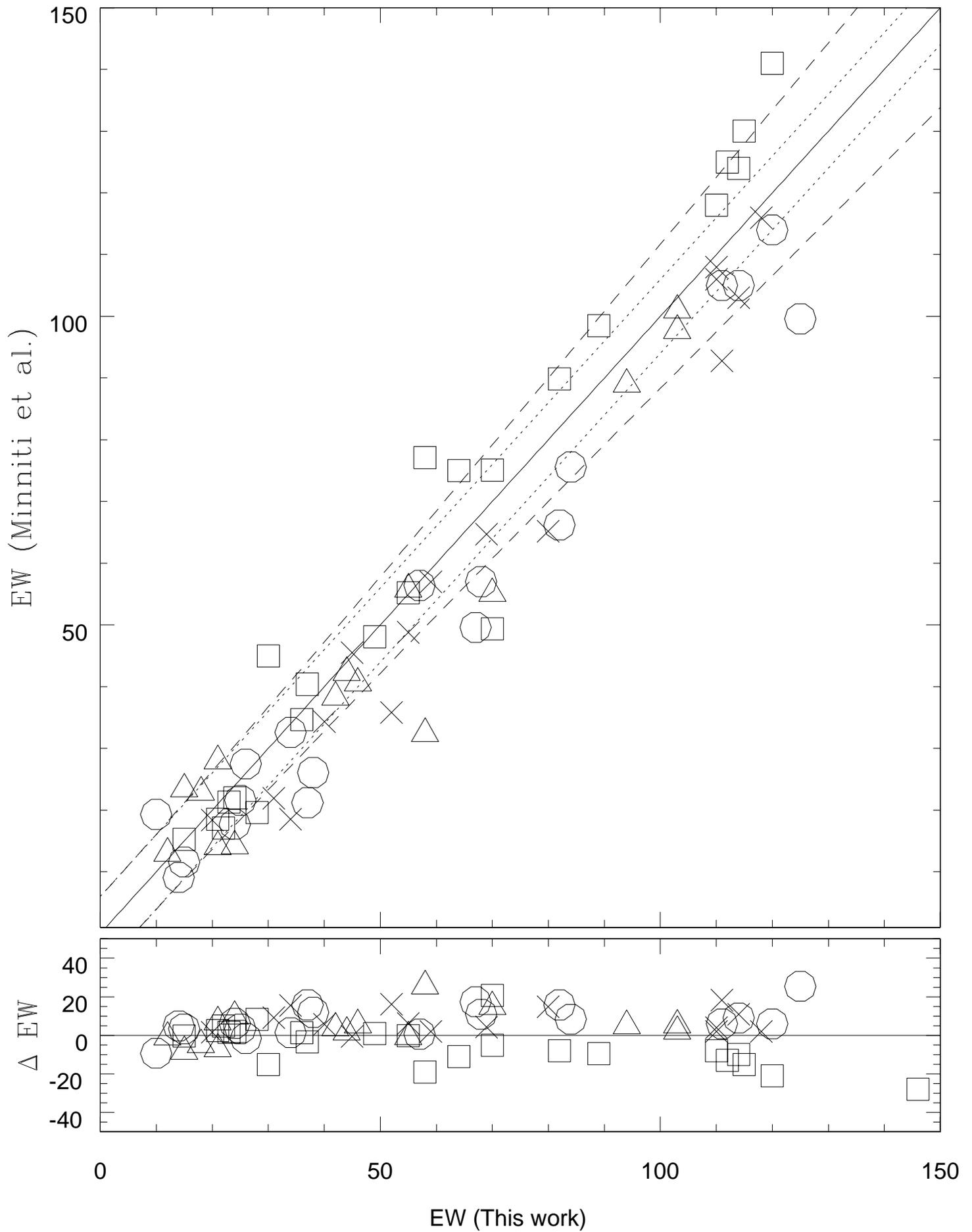
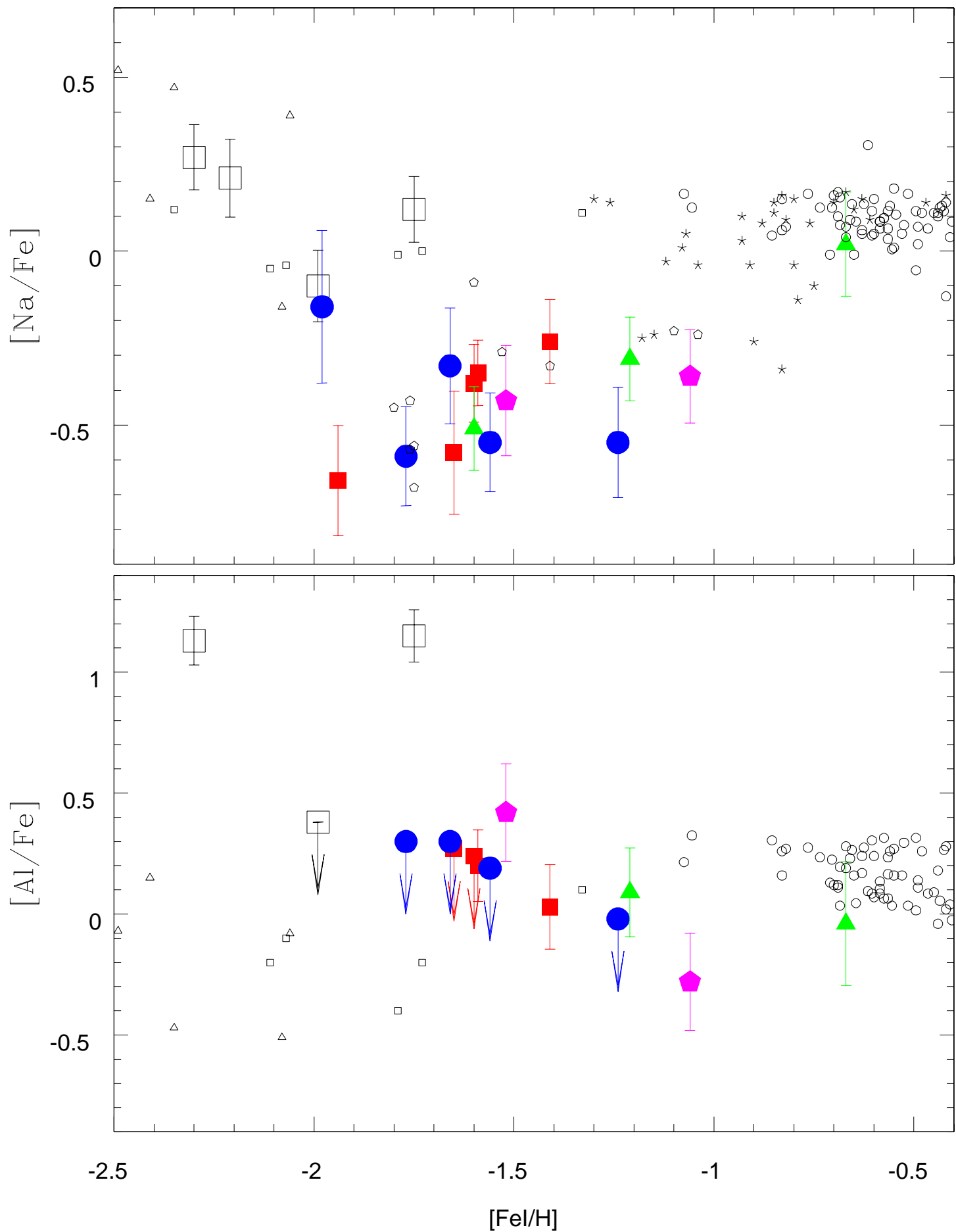
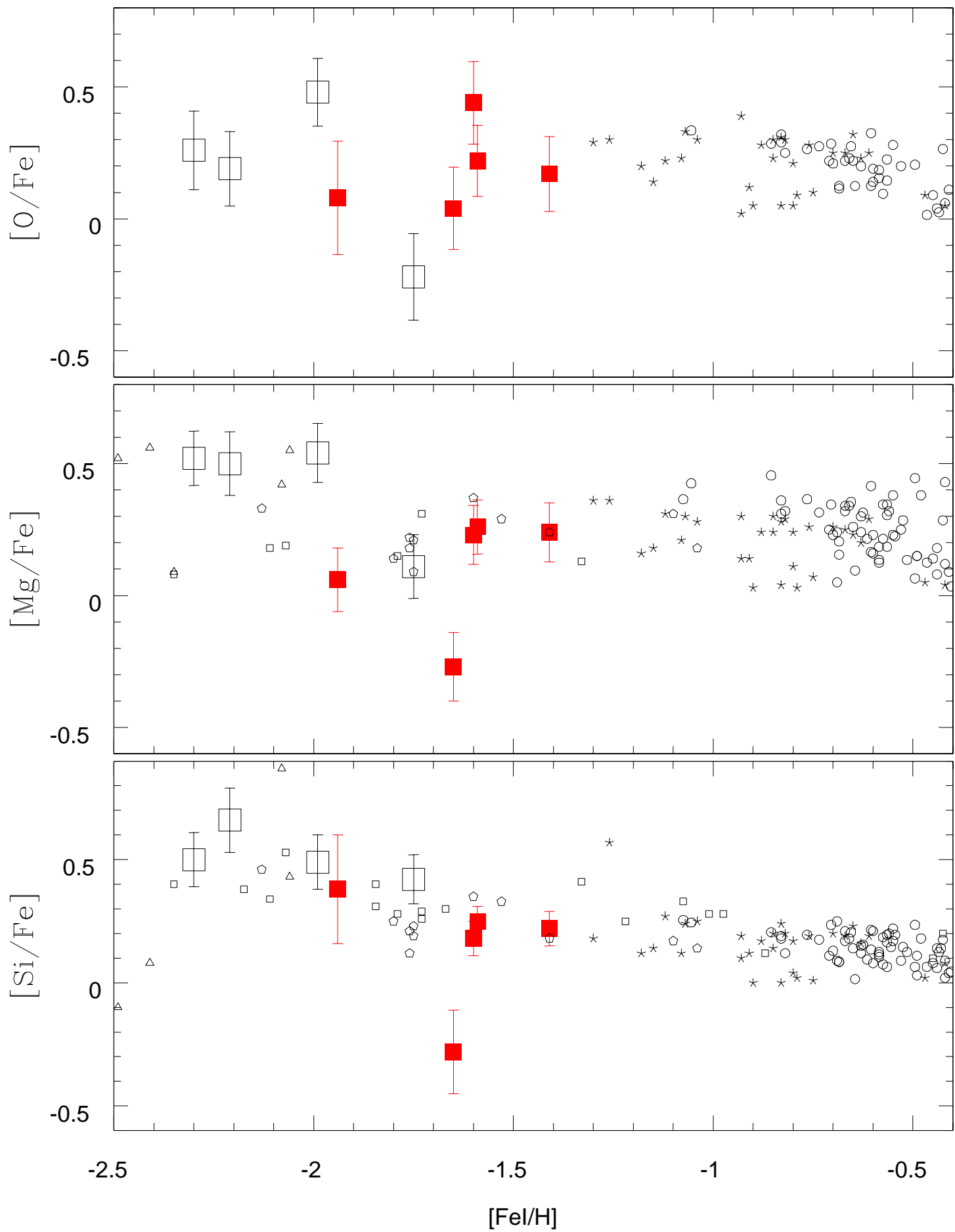
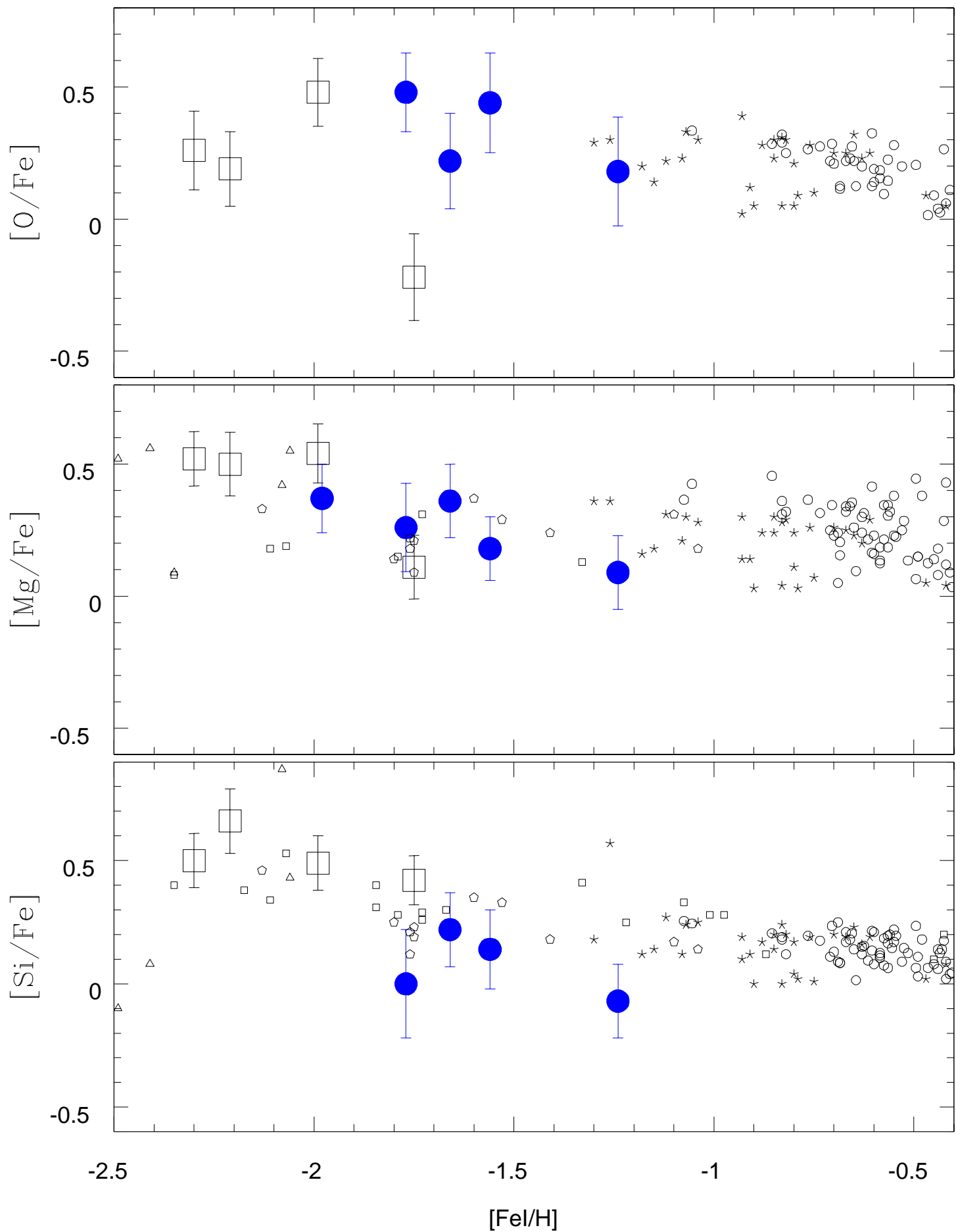
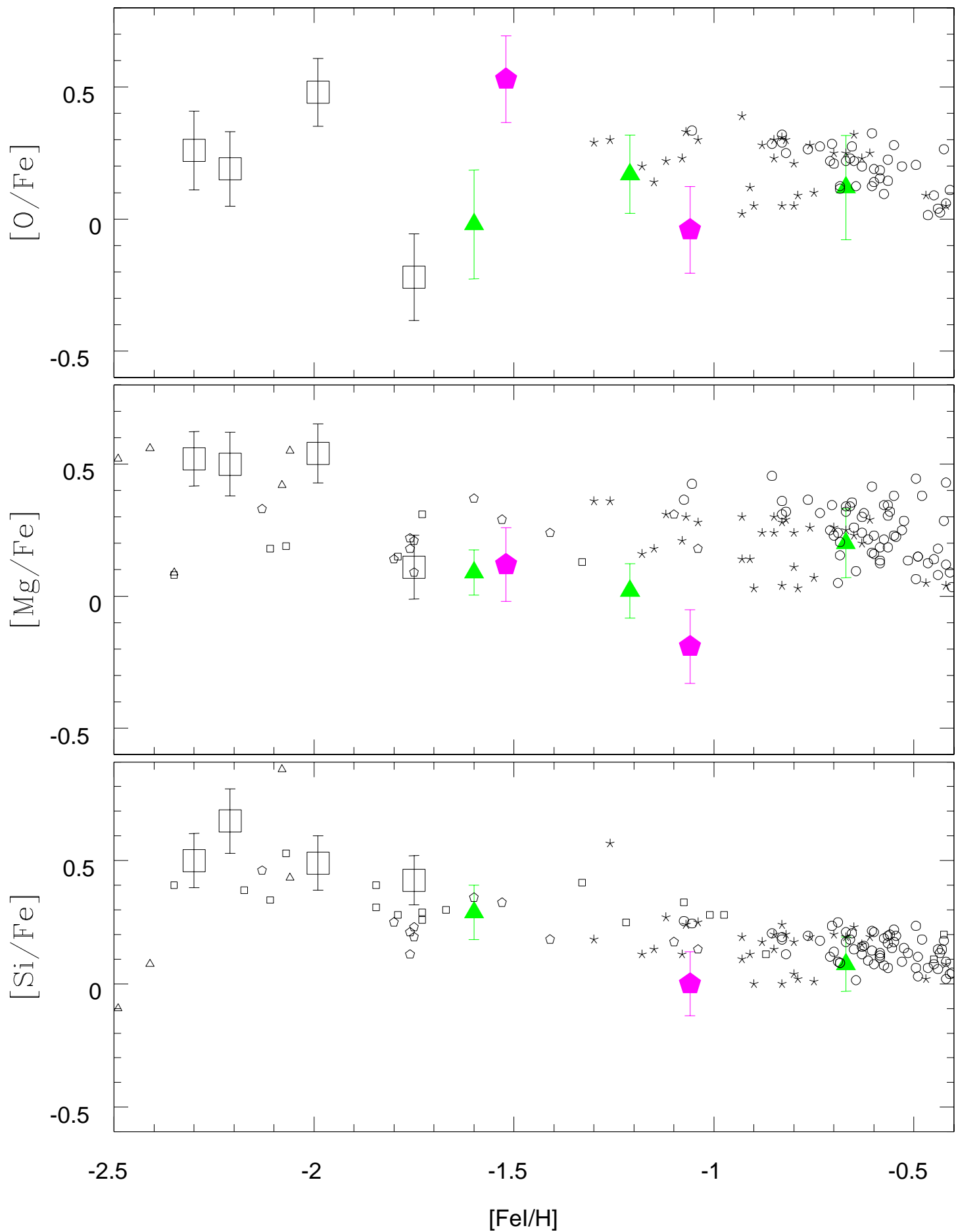
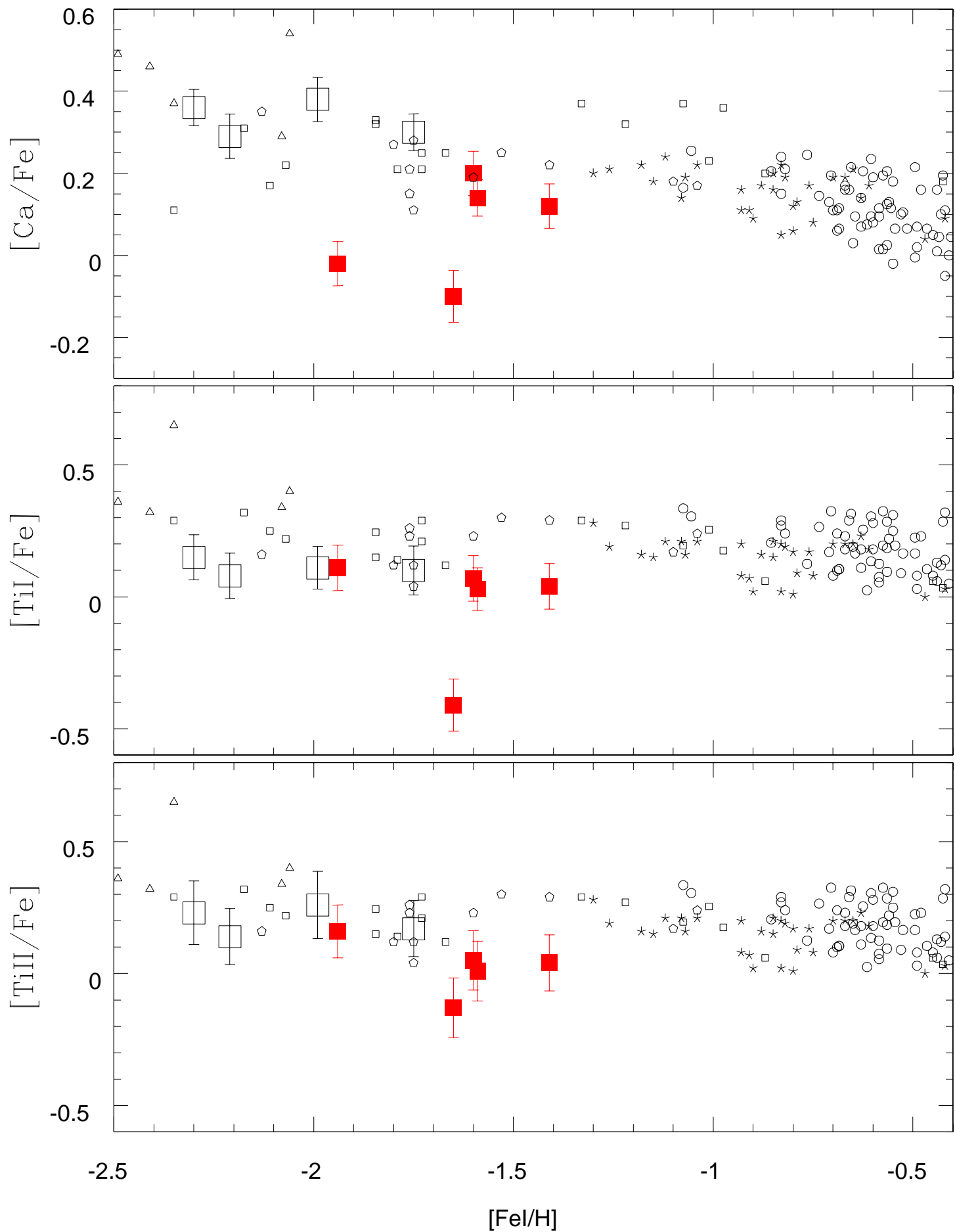
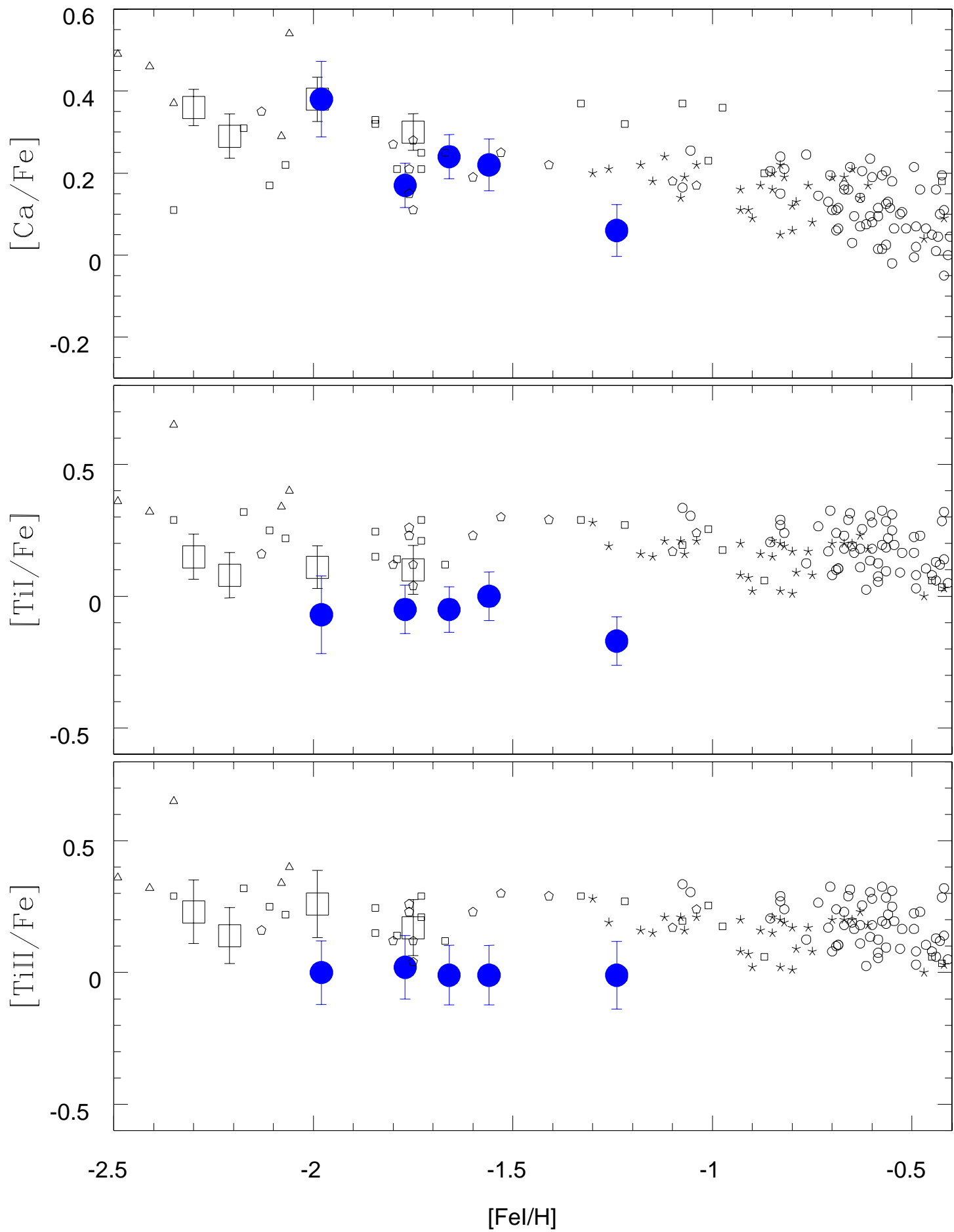
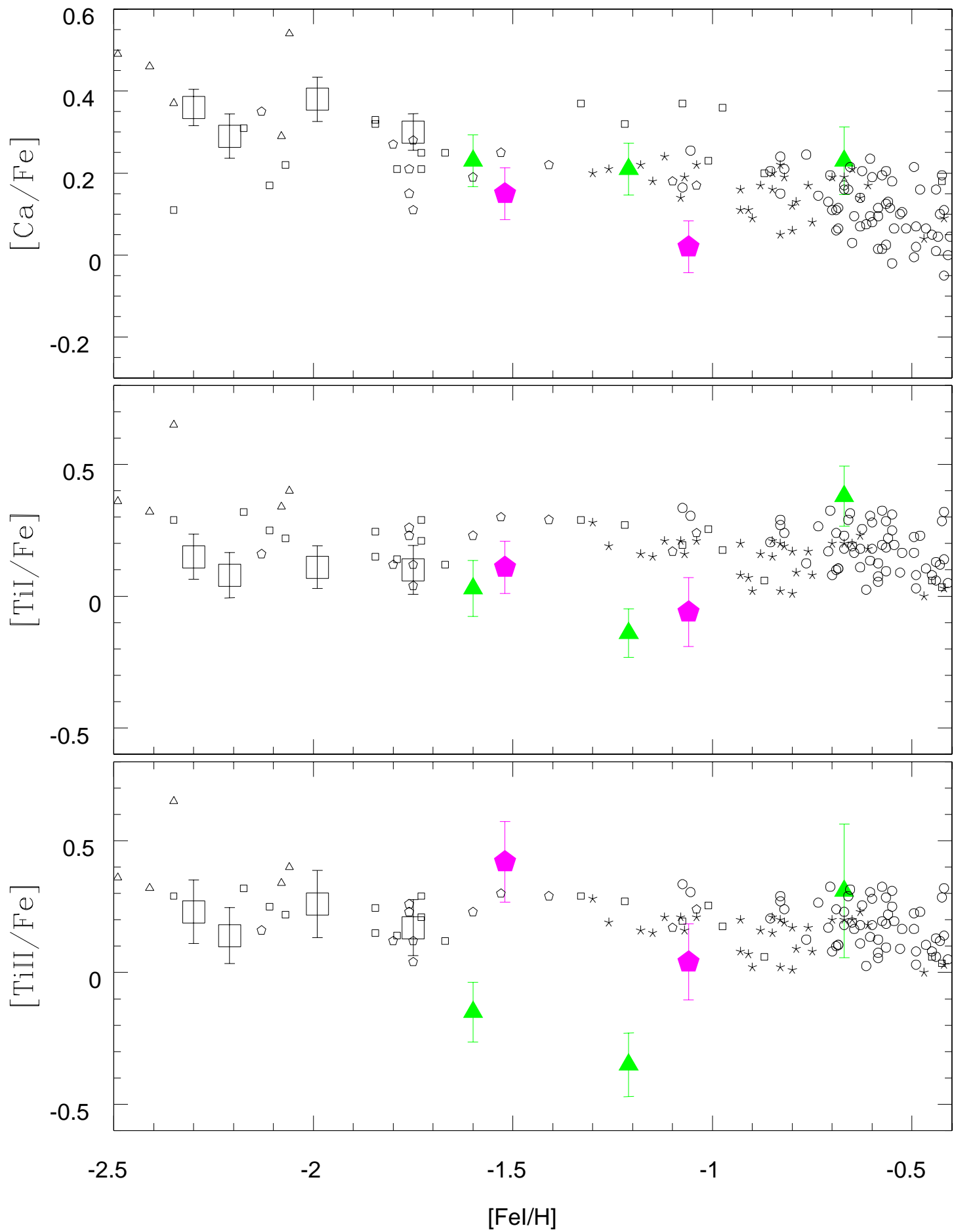
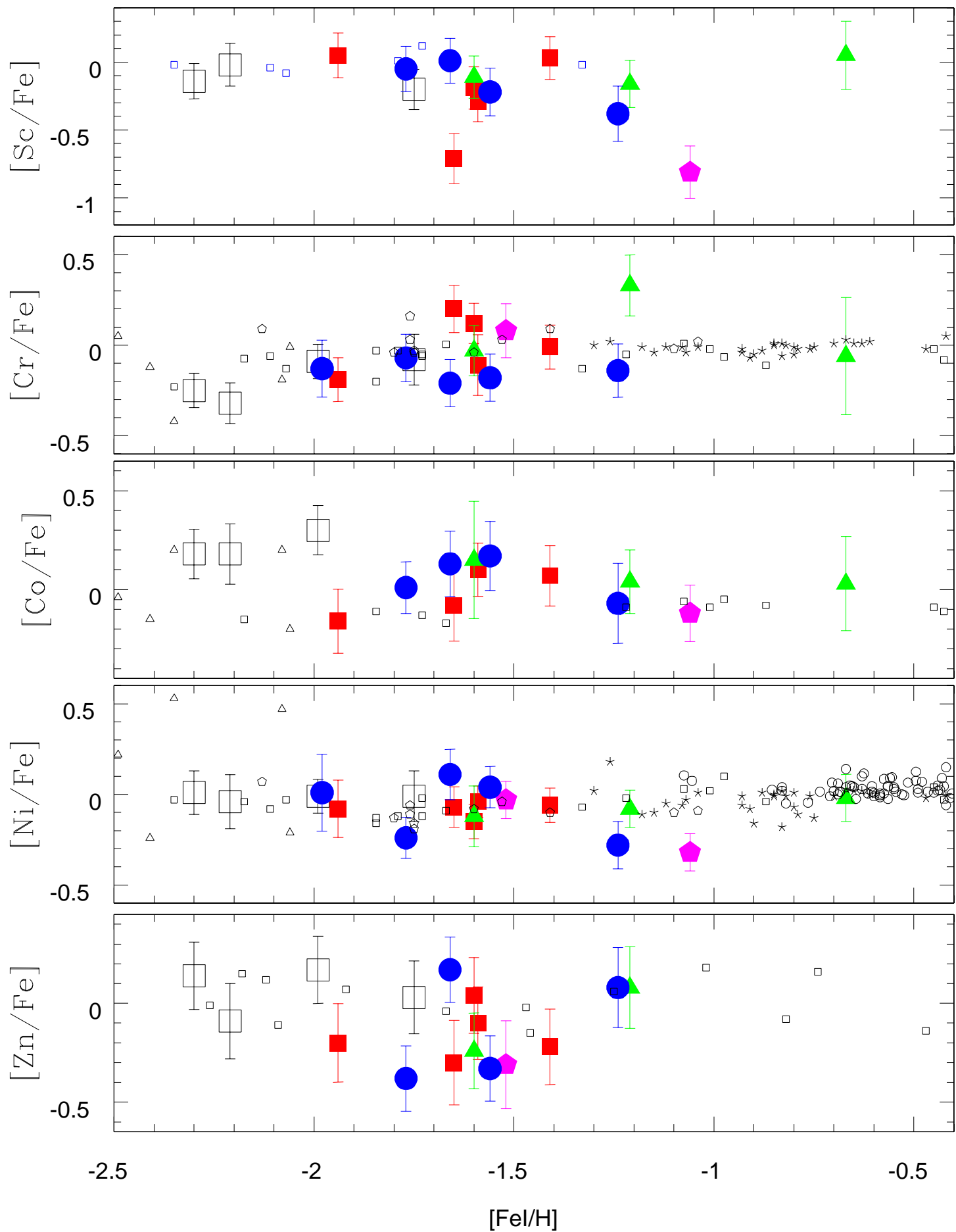
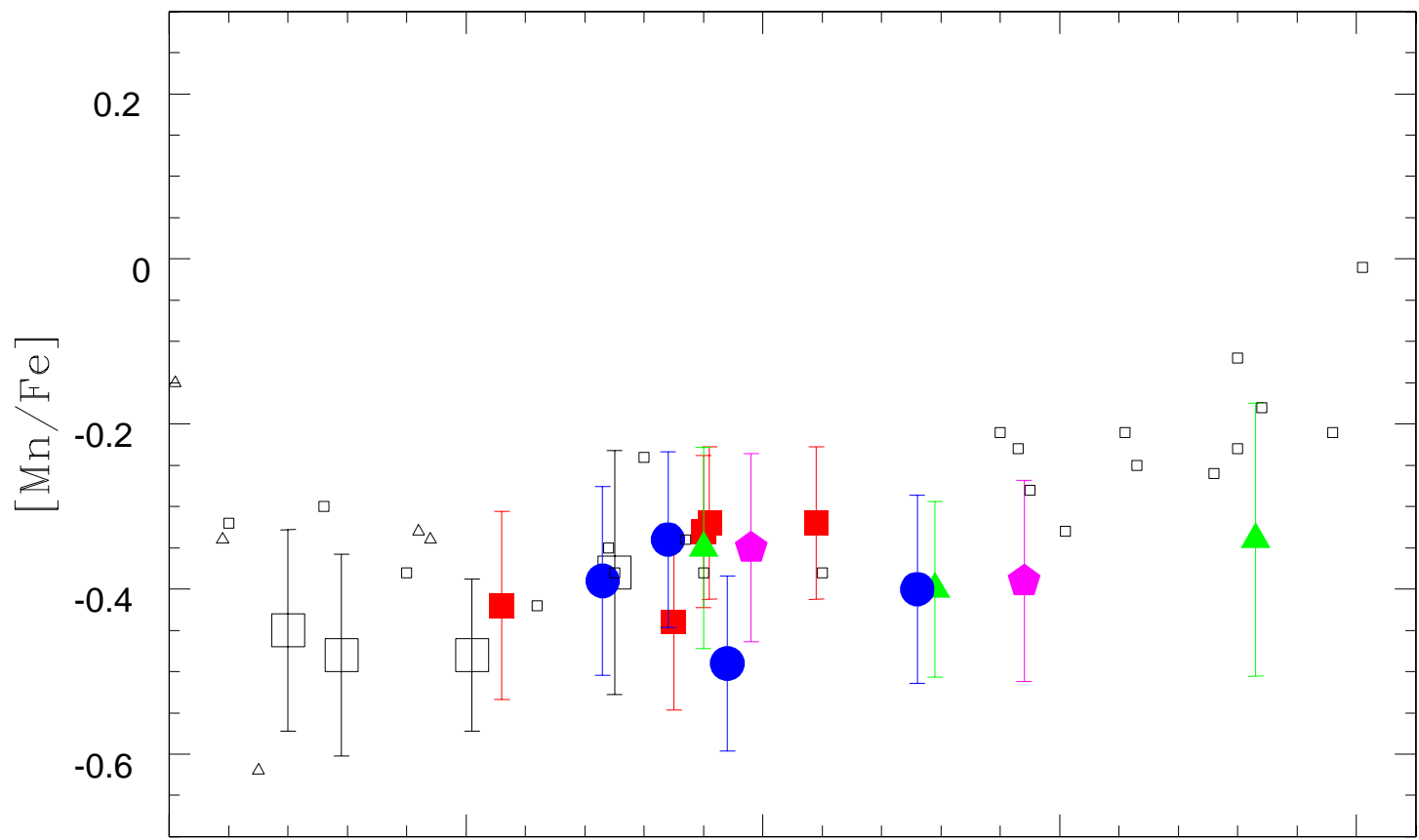
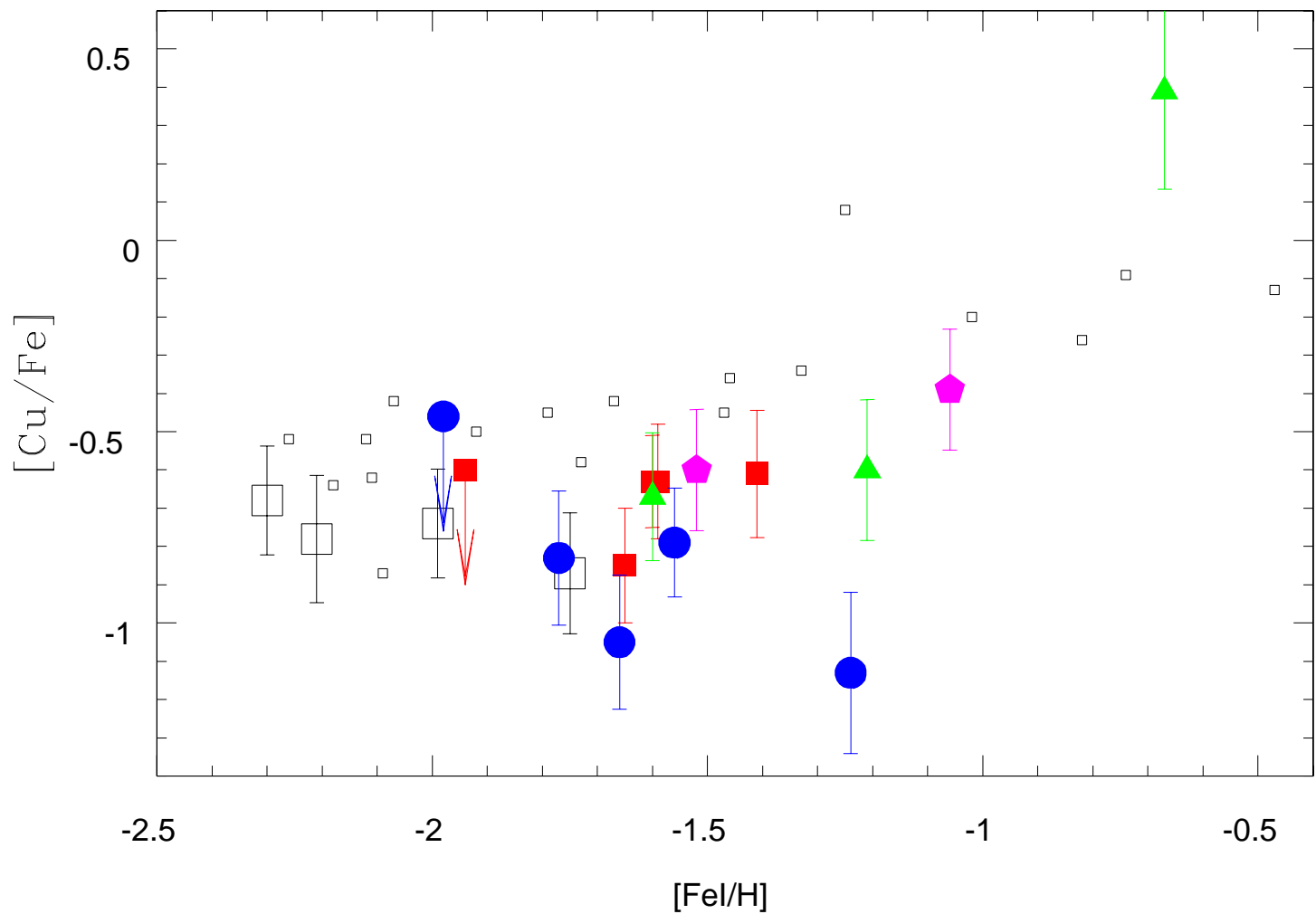
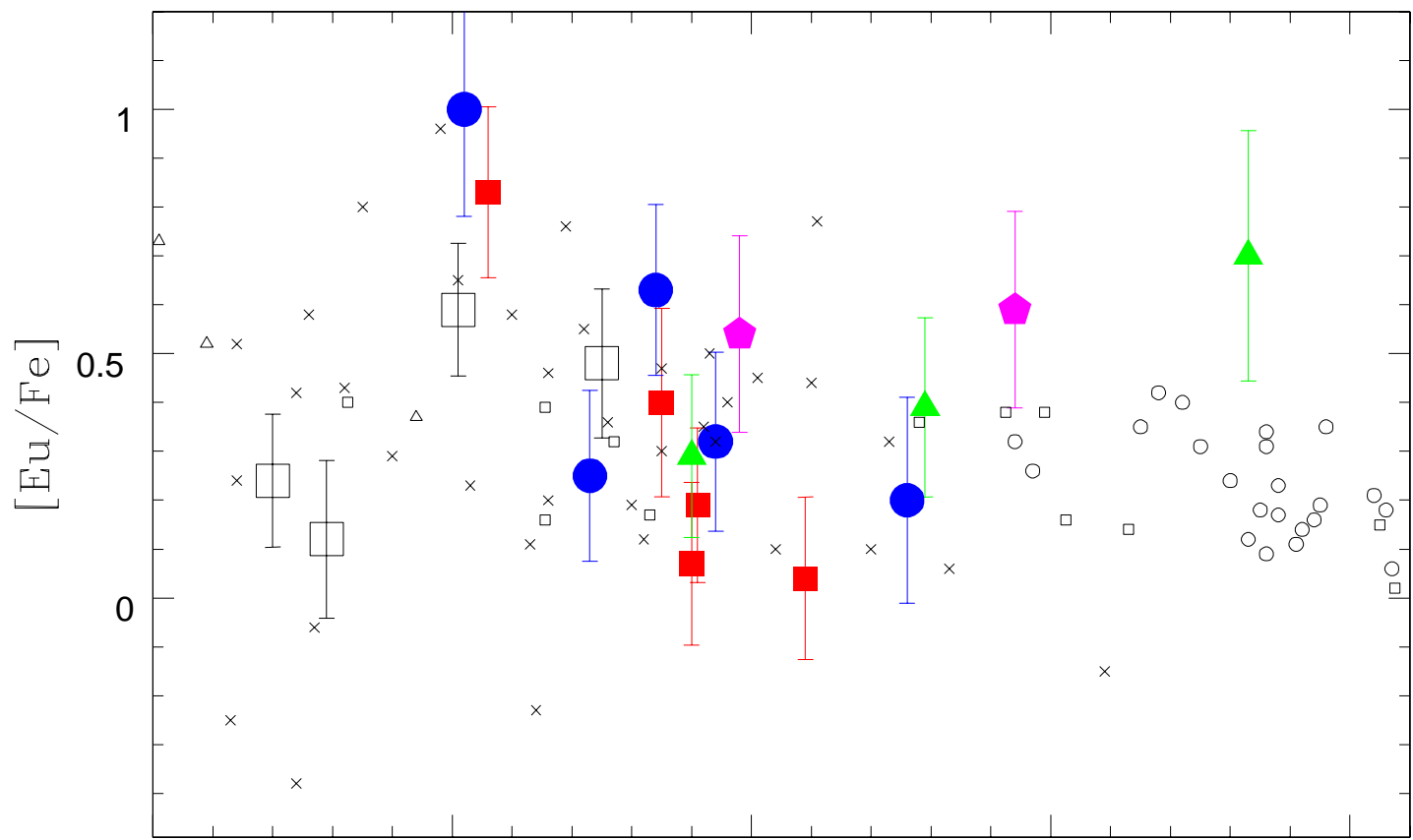
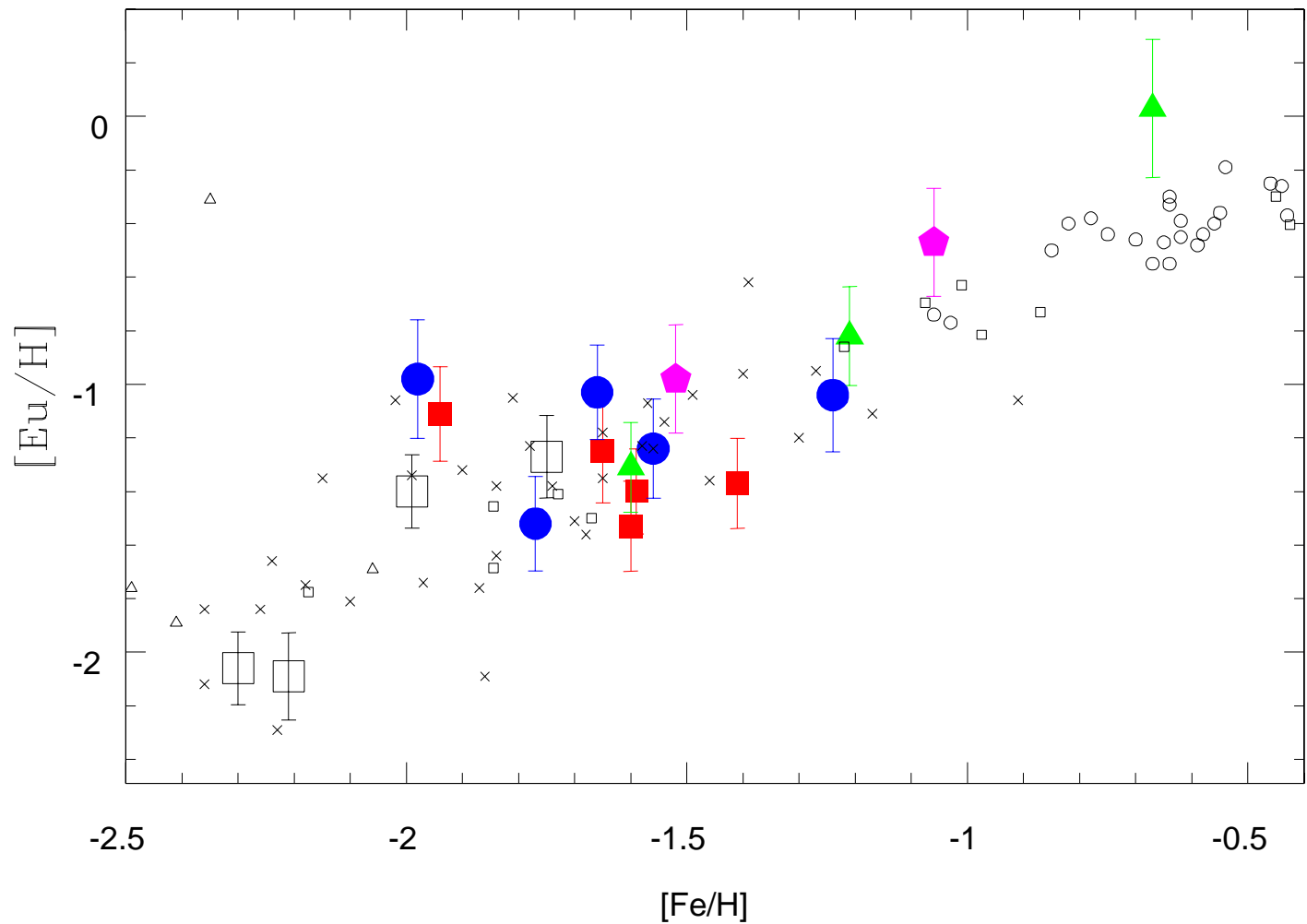
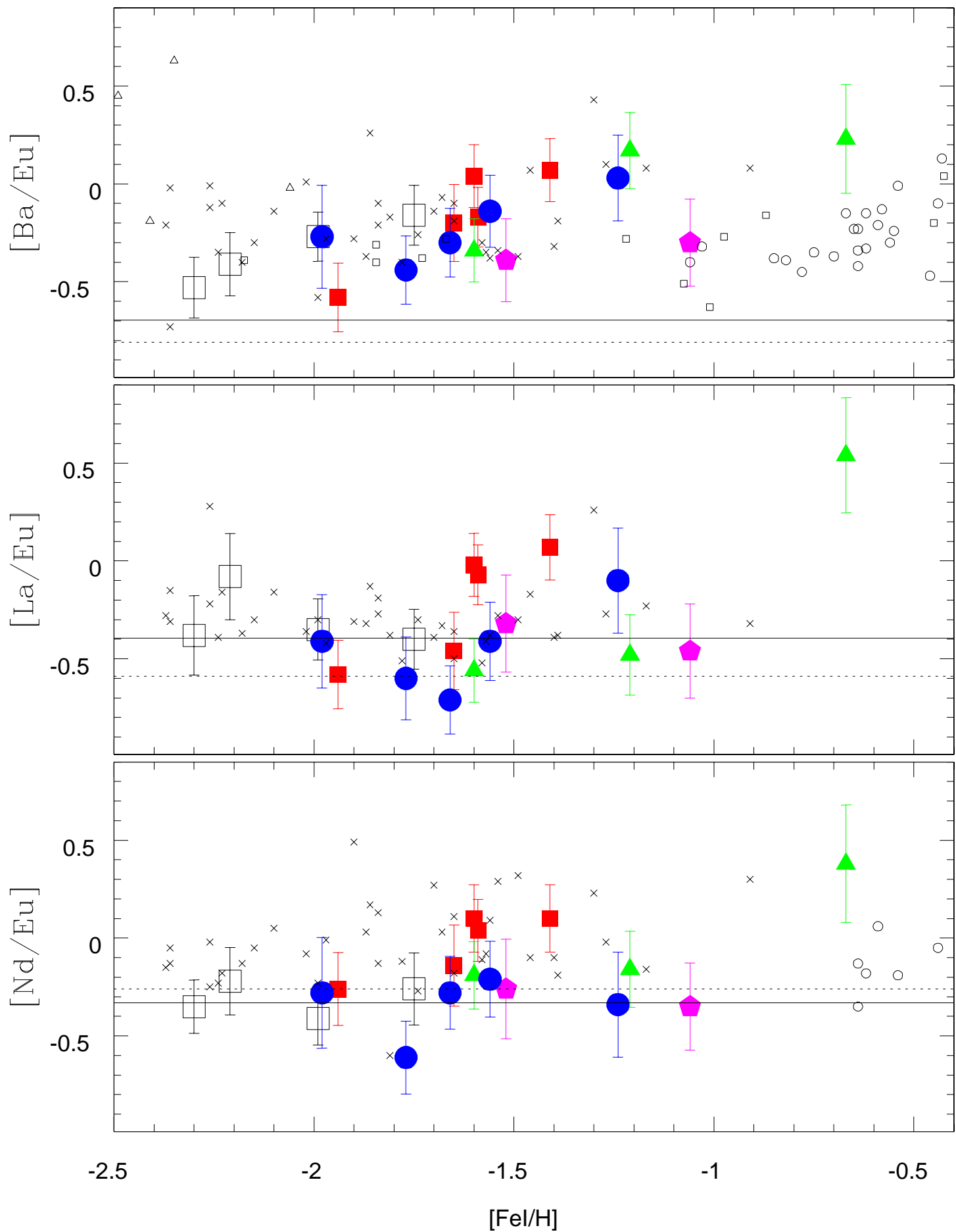
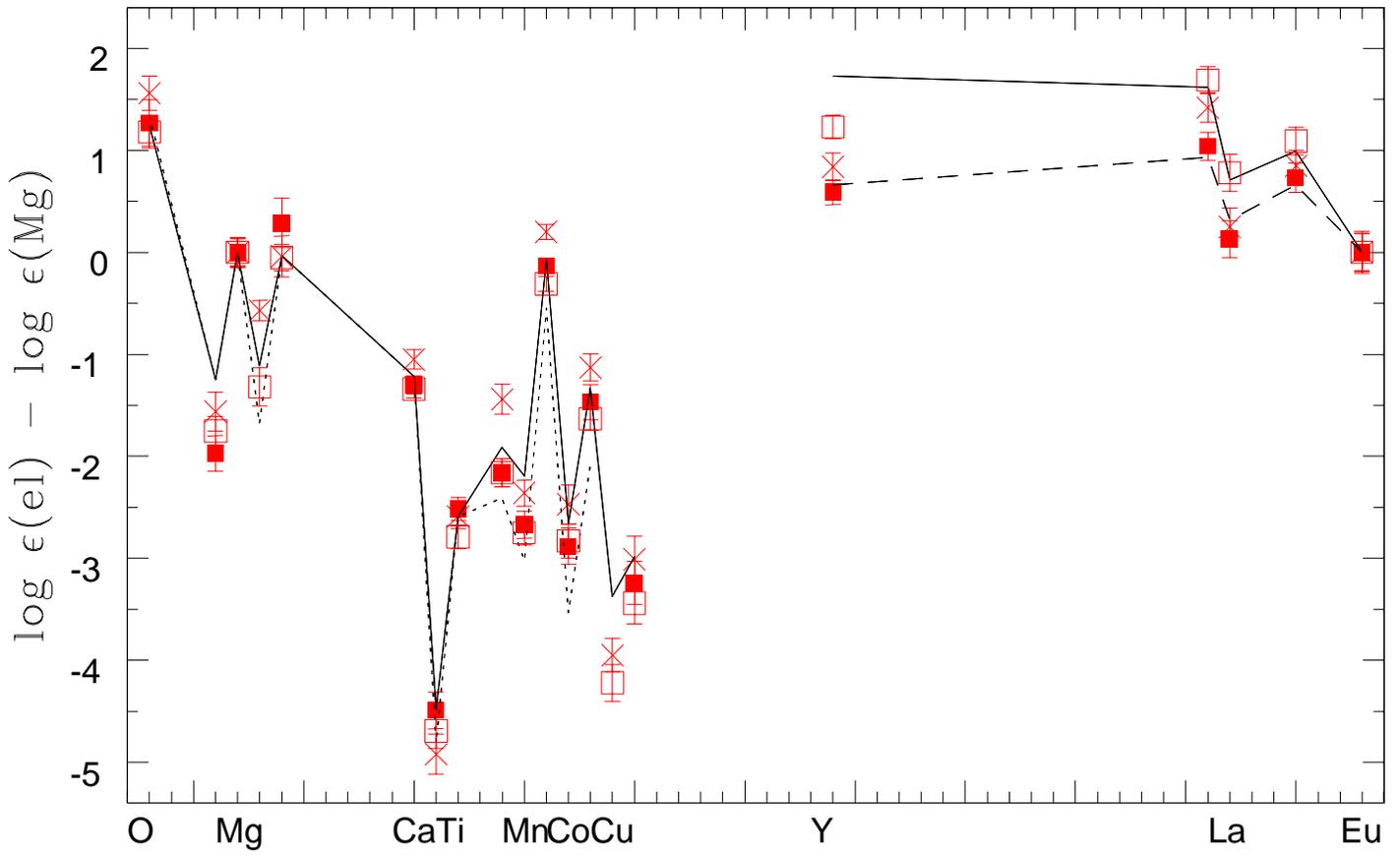
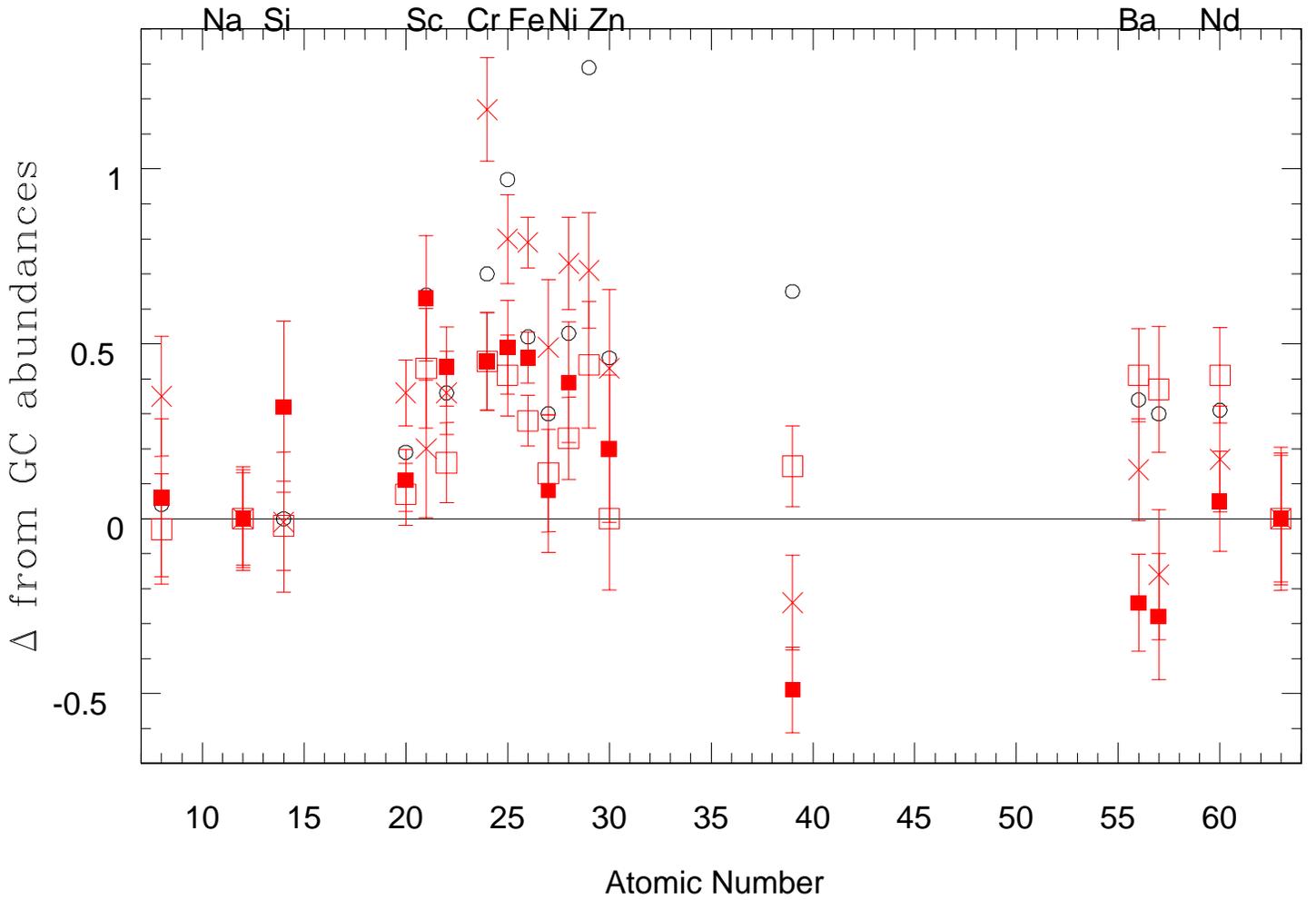
\end{document}